\documentclass[journal]{IEEEtran}
\usepackage{amsmath,amsfonts}
\usepackage{algorithm}
\usepackage{algorithmic}
\usepackage{array}
\usepackage{amssymb} 
\usepackage{textcomp}
\usepackage{stfloats}
\usepackage{url}
\usepackage{booktabs}
\usepackage{tabularx}
\usepackage{verbatim}
\usepackage{graphicx}
\usepackage[caption=false,font=footnotesize]{subfig}
\usepackage{pifont}
\usepackage{cite}
\begin{document}

\title{Beam Hopping Low Earth Orbit Satellite Resource Allocation for Differentiated Services and Robustness Analysis under Model Attacks}

\author{Shuang Zheng, Xing Zhang,~\IEEEmembership{Senior Member,~IEEE,} Quan Z. Sheng, Haixu Wang, \\ and Wenbo Wang,~\IEEEmembership{Senior Member,~IEEE}
        % <-this % stops a space

\thanks{This work is supported by the National Science Foundation of China under Grant 62271062. The work of Shuang Zheng is supported by China Scholarship Council (CSC) under Grant
202406470029. (Corresponding authors: Xing Zhang.)}
\thanks{Shuang Zheng, Xing Zhang, Haixu Wang and Wenbo Wang
are with the School of Information and Communication Engineering, Beijing
University of Posts and Telecommunications, Beijing 100876, China. Shuang Zheng is also with the School of Computing, Macquarie University, Sydney, NSW 2109, Australia. (e-mail:\{zshuang; hszhang; wbwang\}@bupt.edu.cn).}
\thanks{Quan Z. Sheng is with the School of Computing,
 Macquarie University, Sydney, NSW 2109, Australia. (e-mail: michael.sheng@mq.edu.au).}
}

\markboth{Journal of \LaTeX\ Class Files,~Vol.~14, No.~8, April~2026}%
{Shell \MakeLowercase{\textit{et al.}}: A Sample Article Using IEEEtran.cls for IEEE Journals}

\IEEEpubid{0000--0000/00\$00.00~\copyright~2021 IEEE}

\maketitle

\begin{abstract}
Beam hopping (BH)-enabled Low Earth Orbit (LEO) satellites play a pivotal role in next-generation communication networks, providing global coverage, improving spectrum efficiency, and supporting flexible adaptation to heterogeneous service demands. To fully exploit these capabilities, artificial intelligence (AI) techniques are increasingly employed for dynamic resource allocation and power management. However, limited onboard resources and potential adversarial perturbations pose challenges to both efficiency and robustness. To address these issues, we leverage digital twin technology to accurately capture the spatio-temporal dynamics of user–satellite visibility, providing precise state information for decision-making. Building on this, we formulate a joint optimization framework for BH scheduling and power allocation as a Markov Decision Process and propose the BRIDGE—BH with Reinforcement learning incorporating Integrated Dirichlet and Gumbel-TopK Exploration—which integrates a quality of service (QoS)-driven subchannel scheduling mechanism to ensure efficient and differentiated resource allocation. The model’s robustness is systematically evaluated under three classical adversarial attacks. Simulation results demonstrate that our approach achieves superior energy efficiency, service throughput, and fairness, while the robustness analysis shows stable performance under the considered bounded adversarial perturbations.
\end{abstract}


\section{Introduction}
\label{sec1}
\IEEEPARstart{W}{ith} the gradual deployment of 5G, research on 6G technologies has accelerated worldwide to realize the vision of “global coverage, all spectra, full applications, all senses, all digital, and strong security” \cite{1}. Satellite networks represent a paradigm-shifting breakthrough in communications, overcoming the limitations of traditional terrestrial infrastructure and playing a critical role in supporting the development of 6G \cite{40,43}. Among various satellite technologies, Low Earth Orbit (LEO) systems have attracted particular attention from both academia and industry due to their low orbital altitude, reduced deployment costs, and broad coverage compared to Geosynchronous (GEO) and Medium Earth Orbit (MEO) satellites \cite{2,3,4}. For example, the SpaceX Starlink constellation already comprises over 8,000 in-orbit satellites \cite{5}. LEO satellites provide low-latency, high-speed communications and efficiently support differentiated services, establishing a crucial foundation for next-generation communications \cite{6,7}.

Driven by rapidly expanding market demands—the global broadband satellite Internet market is projected to grow from approximately \$8.1 billion in 2025 to \$25.7 billion in 2032 \cite{8}—efficient resource allocation in LEO satellite networks has become increasingly critical. Traditional multi-beam satellites rely on fixed beam directions, often resulting in resource under-utilization and congestion in high-demand regions. In contrast, Beam hopping (BH) technology enables dynamic allocation of onboard power and spectrum, supporting on-demand services while enhancing throughput, spectrum efficiency, and overall service quality \cite{9,10}.

Traditional optimization and heuristic algorithms have been extensively investigated for beam scheduling in satellite communications \cite{11,12,13,14,15,26,27}, which achieve competitive performance. However, the operating environment of LEO satellite systems is inherently dynamic, featuring rapid satellite mobility, continuously evolving user–satellite visibility, and highly time-varying traffic demands. Under such conditions, traditional optimization and heuristic approaches typically require frequent iterative re-optimization to adapt to system states, which incurs substantial computational overhead. As a result, these methods exhibit fundamental limitations in tracking time-varying dynamics and satisfying real-time decision-making requirements, particularly in light of the stringent onboard computational and energy constraints of LEO satellites \cite{42}. To overcome these challenges, artificial intelligence (AI) and machine learning techniques, particularly deep reinforcement learning (DRL), have emerged as promising tools for intelligent resource management \cite{16,17,18,19,20,28,29,30}.
Moreover, AI-driven scheduling models are susceptible to robustness and security challenges, including adversarial attacks and abnormal input perturbations. Thus, resource allocation in BH LEO satellites still faces the following core challenges:
\IEEEpubidadjcol
\begin{itemize}
\item{\textit{Differentiated Service Transmission with Limited On-board Resources}: LEO satellite networks support diverse services with heterogeneous quality of service (QoS) requirements. Given the limited onboard power and spectrum, LEO satellites must implement efficient and fair resource allocation, while ensuring that high-priority services are served preferentially to maximize overall resource utilization and ensure reliable, high-quality communications.}

\item{\textit{Dynamic LEO User-Satellite Connections}: The continuous motion of LEO satellites induces significant spatio-temporal variations in the communication links with ground users. Channel quality depends on relative positions, propagation distances, and environmental factors, while satellite visibility is inherently time-varying. These dynamics increase the complexity of network management and BH scheduling, necessitating real-time link state awareness and adaptive beam and resource allocation strategies to maintain performance in dynamic network scenarios.}

\item{\textit{Robustness and Model Security in AI Architectures}: AI-based resource allocation models are susceptible to robustness and security challenges, including adversarial attacks and abnormal input perturbations. Systematic research on the robustness of DRL-based resource scheduling models for LEO satellite networks under such threats remains limited, highlighting the need for further studies to ensure performance and trustworthiness in next-generation communication networks.}
\end{itemize}

To address these challenges, we propose a joint optimization framework that formulates the BH scheduling and power allocation problem in LEO satellite systems as a Markov decision process (MDP), explicitly accounting for the time-varying user–satellite dynamics and heterogeneous QoS requirements. Based on the formulation, we design an AI-enabled algorithm to efficiently manage the hybrid decision space, improve resource utilization, and analyze robustness under adversarial conditions. The main contributions of this work are summarized as follows:
\begin{itemize}
\item We 
%introduce 
exploit the 
digital twin (DT) technology to accurately capture the spatio-temporal dynamics of user–satellite visibility, providing high-fidelity state information that supports adaptive and efficient BH scheduling.

\item We propose the BH with Reinforcement learning incorporating Integrated Dirichlet and Gumbel-TopK Exploration (BRIDGE) algorithm, which jointly optimizes BH scheduling and power allocation in discrete–continuous action spaces, while embedding a QoS-driven subchannel mechanism to ensure differentiated service provisioning.

\item We systematically assess the vulnerability of DRL-based scheduling models to classical adversarial attacks, including the Fast Gradient Sign Method (FGSM), Iterative FGSM (I-FGSM), and Projected Gradient Descent (PGD), thereby providing a robustness evaluation for AI-enabled BH LEO satellite systems.

\item Extensive simulations demonstrate that the proposed framework outperforms baseline methods in energy efficiency, service throughput, and fairness. The robustness analysis further shows that it maintains stable performance under the considered bounded adversarial perturbations.
\end{itemize}

The remainder of this article is organized as follows. Section \ref{sec2} reviews the related work. Section \ref{sec3} presents the system model and problem formulation. Section \ref{sec4} introduces the proposed BRIDGE algorithm. Section \ref{sec5} analyzes the adversarial attack model and the corresponding algorithm. Section \ref{sec6} evaluates the performance of the proposed algorithm in comparison 
%with 
to 
existing baseline methods. Finally, Section \ref{sec7} concludes the paper.

\section{Related Work}
\label{sec2}
Research on BH resource allocation in LEO satellite networks can be broadly classified into traditional optimization-based approaches and AI-enabled methods. With the increasing integration of AI into communication systems, concerns have also emerged regarding the robustness and security of intelligent scheduling models.
%have also emerged. 
This section reviews the related studies from these perspectives.

\subsection{BH Scheduling and Resource Allocation based on Traditional Optimization  Algorithm}
Traditional optimization-based approaches for BH scheduling typically rely on heuristic methods or mathematical programming to achieve multi-dimensional resource allocation under diverse traffic demands. For example, \cite{11,12} applied greedy strategies for beam scheduling, where \cite{11} further incorporated isolation distance constraints to mitigate inter-beam interference, while \cite{12} employed a quadratic transformation to optimize power allocation under a given BH pattern, thereby improving service satisfaction and throughput. Similarly, \cite{13} decomposed the original mixed-integer non-convex problem into three subproblems—beam direction control with timeslot allocation, subchannel assignment, and power allocation—which were solved iteratively via an externality-matching scheme combined with the successive convex approximation (SCA) method.

Beyond these optimization frameworks, heuristic algorithms have also been widely explored \cite{14,15,26,27}. For instance, \cite{14} compared genetic algorithms, particle swarm optimization, and simulated annealing for coverage optimization and interference suppression. Meanwhile, \cite{15,26,27} employed genetic algorithms for multi-objective optimization of resource allocation, targeting different performance metrics such as throughput, access success rate, latency, and fairness. These approaches demonstrated notable advantages, particularly in scenarios with uneven traffic demands.
\begin{table*}[t]
\centering
\caption{Comparison of AI-enabled BH resource allocation studies and the proposed BRIDGE framework.}
\label{tab:related_work_comparison}
\scriptsize
\renewcommand{\arraystretch}{1.25}
\setlength{\tabcolsep}{2.8pt}
\begin{tabularx}{\textwidth}{
p{1.2cm}
X
X
X
X
X
X
X}
\toprule
\textbf{Studies} 
& \textbf{Method Type} 
& \textbf{Beam Scheduling} 
& \textbf{Power Allocation} 
& \textbf{Hybrid Action Handling} 
& \textbf{User--Satellite Visibility} 
& \textbf{Differentiated Services} 
& \textbf{Robustness} \\
\midrule

{\cite{16}, \cite{17}} 
& Hybrid-action DQN
& Discrete beam scheduling based on value-function approximation 
& Continuous power allocation modeled as action parameters 
& Parameterized value-based hybrid action representation 
& Not explicitly modeled 
& Limited consideration 
& No \\
\cite{29,18} 
& PPO/multi-agent PPO 
& Policy-gradient-based resource scheduling 
& Power allocation by policy learning 
& Hybrid policy learning
& Not explicitly modeled 
& Limited consideration 
& No \\
\cite{19,20} 
& Multi-objective DRL 
& DRL-based beam scheduling 
& Decoupled or convex-transformed power allocation 
& Not a fully joint discrete--continuous policy 
& Not explicitly modeled 
& RT/NRT services considered 
& No \\

\textbf{BRIDGE} 
& \textbf{PPO-based hybrid-action DRL} 
& \textbf{Gumbel-TopK Sampling} 
& \textbf{Dirichlet Sampling} 
& \textbf{Joint discrete-continuous policy } 
& \textbf{DT-based user-satellite visibility windows} 
& \textbf{QoS-aware subchannel allocation for RT/NRT services} 
& \textbf{FGSM, I-FGSM, PGD} \\
\bottomrule
\end{tabularx}
\label{tab_0}
\end{table*}
\subsection{AI-Enabled Beam Scheduling and Power Allocation Optimization}
In BH-enabled LEO satellite networks, traditional optimization and heuristic methods face limitations under highly dynamic traffic conditions. To address these challenges, DRL has been increasingly applied to resource allocation, with several approaches treating beam scheduling and power allocation as decoupled subproblems. The authors in \cite{28} optimized discrete beam selection using DRL while handling power allocation independently to reduce computational complexity. In \cite{20}, beam scheduling was guided by traffic demands, with power adjusted separately to enhance responsiveness under dynamic conditions. \cite{30} considered multi-satellite non-orthogonal multiple access (NOMA) systems, where beamforming and subchannel assignment were optimized prior to independent power control to improve overall system efficiency. Although these decoupled strategies simplify problem-solving, they do not fully exploit the coupling between beam scheduling and power allocation, limiting overall performance and service quality. 

To address these limitations, \cite{16} and \cite{17} introduced hybrid-action deep Q-network (DQN) frameworks that integrate discrete beam scheduling with continuous power allocation via a parameterized action space. Notably, \cite{17} extended this approach to jointly optimize multiple objectives, including transmission delay, packet loss, and power consumption, demonstrating the potential of hybrid-action learning in complex satellite communication scenarios. However, since P-DQN relies on value function approximation, its exploration capability is limited, and it may still converge to local optima in highly complex or dynamic scenarios. More recently, policy gradient methods, such as Proximal Policy Optimization (PPO), have been introduced to improve exploration and learning stability \cite{29,18}. A two-stage optimization framework combining multi-agent PPO was proposed in \cite{18} to jointly optimize frequency planning and beam power allocation in high-throughput satellite systems. 

To support diverse service types in LEO satellites, \cite{19} and \cite{20} optimized multiple performance metrics for simultaneous real-time (RT) and non-real-time (NRT) transmissions. Specifically, \cite{19} employed a multi-objective DRL scheme for BH scheduling in multi-beam satellites, while our previous work \cite{20} introduced an action-masking multi-objective double deep Q-network (AMM-DDQN) to enable flexible beam scheduling and transform the power optimization problem into a convex form for dynamic traffic and channel conditions.

\subsection{Security and Robustness of AI-Enabled Scheduling Model}
Prior studies on general AI models have revealed their high vulnerability to adversarial attacks, which can lead to substantial performance degradation \cite{31,32,33}. In particular, DRL agents used for decision-making in complex environments are especially susceptible. For example, \cite{21} compared the effects of random noise and adversarial samples generated by the Fast Gradient Sign Method (FGSM) on DRL agents, demonstrating that FGSM samples are significantly more effective at misleading the agents than random perturbations. Similarly, \cite{22} examined multiple DRL algorithms—including PPO, Deep Deterministic Policy Gradient (DDPG), and DQN—and reported pronounced performance drops under PGD attacks.

%Y.A 
Ergu et al. \cite{23,24,25} conducted a series of studies on DRL-based open radio access network (O-RAN) resource allocation. The work in \cite{23} revealed vulnerabilities of DRL-enabled resource management strategies to FGSM, PGD, and Policy-Induced Attacks (PIA). Subsequent studies \cite{24,25} extended this analysis to dynamic vehicle-to-everything (V2X) scenarios and proposed robust frameworks, such as targeted RADAR, to enhance the stability and reliability of DRL systems under adversarial conditions.

Although existing studies have shown that DRL-based methods can flexibly adapt to the spatio-temporal dynamics of traffic demands and perform dynamic beam scheduling, the modeling of onboard total power constraints remains rigid, and the dynamic connectivity between users and satellites is often overlooked. Besides, existing hybrid-action frameworks are fundamentally constrained in large-scale LEO BH scenarios, where scheduling requires selecting beam subsets from a large candidate set, leading to a high-dimensional combinatorial action space. In such cases, value-based methods relying on discrete enumeration or parameterized actions face inherent scalability and exploration challenges, which significantly limits their applicability to practical large-scale beam scheduling. Furthermore, subchannel allocation among heterogeneous users within a beam has not been comprehensively addressed. Thus, we incorporate user–satellite visibility into a joint beam scheduling and power allocation optimization framework for LEO satellite systems, specifically targeting differentiated user requirements. In addition, the robustness of the proposed model is systematically evaluated under adversarial attack scenarios to ensure reliable and resilient performance in dynamic environments. To further clarify the distinctions between the proposed BRIDGE framework and existing AI-enabled studies, Table \ref{tab_0} provides a comparative summary from multiple perspectives.

\begin{figure}[!t]
\centering
\includegraphics[width=\columnwidth]{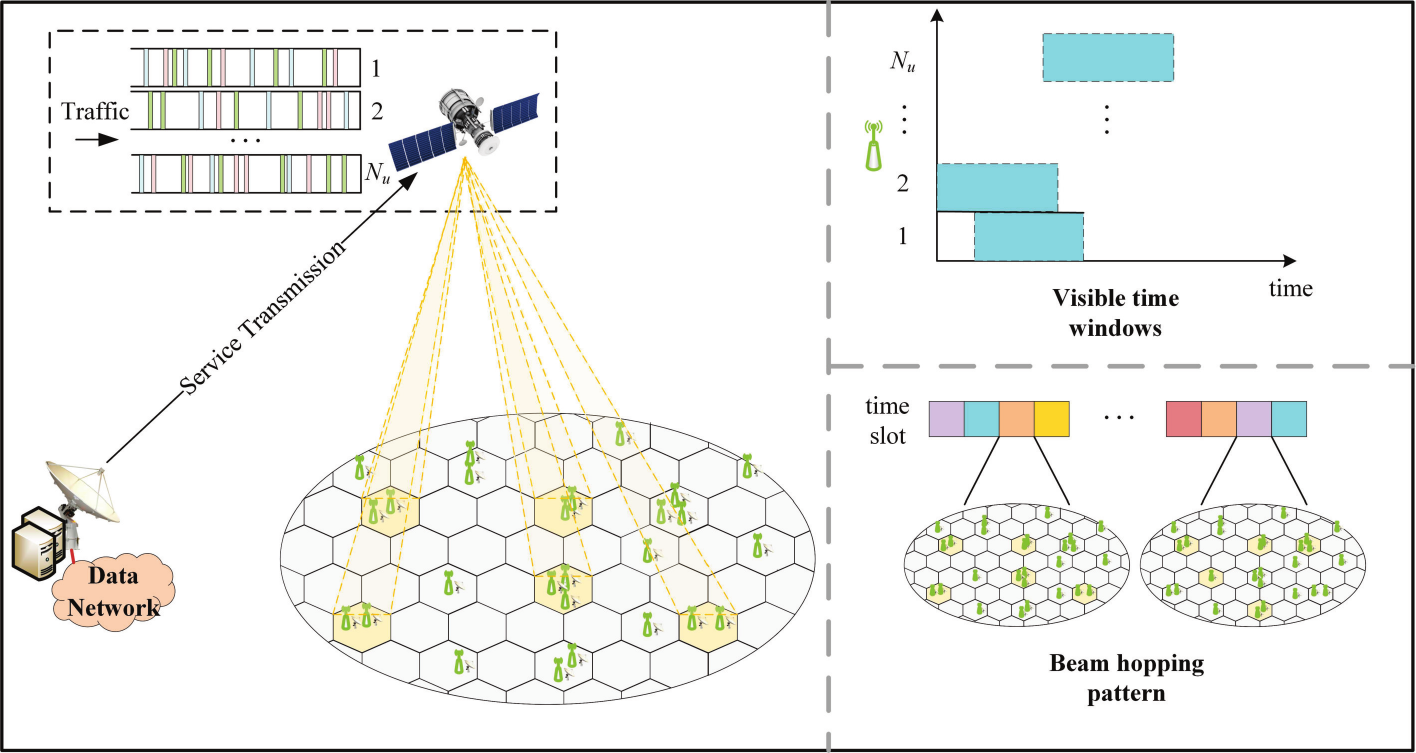}
\vspace{-4mm}
\caption{Downlink transmission in the BH LEO satellite system.}
\label{fig_1}
\end{figure} 

\section{System Model and Problem Formulation}
\label{sec3}
This section provides a comprehensive overview of the downlink operation in a BH LEO satellite network. We begin by describing the network architecture and the key components of the satellite–user communication links. Next, we present the traffic and communication models that capture the spatio-temporal dynamics of user demand and channel conditions. Finally, we formulate the joint beam scheduling and power allocation problem, which aims to optimize multiple performance metrics while accounting for system constraints and user differentiation.

\subsection{Network Architecture}
As illustrated in Fig.~\ref{fig_1}, we consider the downlink transmission scenario, where an LEO satellite serves a set of ground users $\mathcal{U}=\{1,2,\ldots ,{{N}_{u}}\}$, and the corresponding beam positions within the coverage area are represented by $\mathcal{B}=\{1,2,\ldots ,{{N}_{b}}\}$. Over the discrete time slot set $\mathcal{T}=\{1,2,\ldots ,T\}]$, the user-beam associations are denoted by $\mathcal{A}=\{a_{u,b}^{t}|u\in \mathcal{U},b\in \mathcal{B},T\in \mathcal{T},a_{u,b}^{t}=\{0,1\}\}$. The LEO satellite is equipped with a phased-array antenna that supports BH, allowing at most $K$ co-channel beams to be simultaneously activated in each BH slot, covering a subset of beam positions $\mathcal{K}^t$. A binary variable $y_{k,u}^{t}$ indicates whether the $u$-th user is served by the $k$-th beam during time slot $t$, with $y_{k,u}^{t}=1$ if served and $y_{k,u}^{t}=0$ otherwise.

Due to the continuous motion of LEO satellites, ground users maintain line-of-sight (LOS) connectivity with a satellite only within specific temporal intervals. Let $l_{u,s}(t)$ and $\mathcal{S}_o$ denote the direct propagation path between user $u$ and satellite $s$ at time $t$ and the set of geometric objects in the 3D environment, respectively. The LOS visibility indicator is defined as:
\begin{equation}
v_{u,s}(t) =
\begin{cases}
1, & \text{if } \ell_{u,s}(t) \cap \mathcal{S} = \varnothing, \\
0, & \text{otherwise}.
\end{cases}
\end{equation}

\begin{figure}[!t]
\centering
\includegraphics[width=\columnwidth]{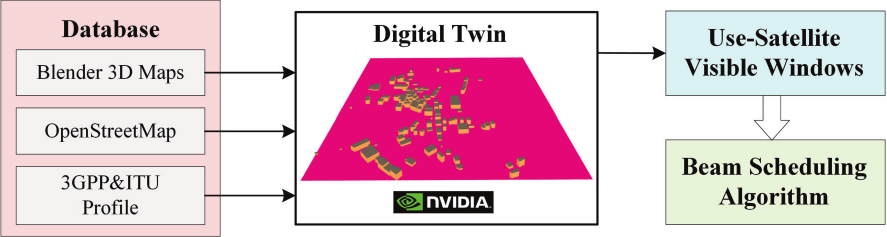}
\vspace{-4mm}
\caption{Digital twin–based modeling of user–satellite visibility windows.}
\label{fig_2}
\end{figure}

Accordingly, the visibility window of the $u$-th user is defined as $\mathcal{T}_{u,s} \subseteq \mathcal{T}$, which consists of the time slots when the satellite $s$ satisfies $v_{u,s}(t) = 1$. User–satellite visibility windows are generated using NVIDIA Sionna, a GPU-accelerated, open-source link-level simulator that incorporates a physics-based ray tracer built on Mitsuba 3 and TensorFlow \cite{34}. In this work, Sionna is employed to determine the existence of a direct LOS path between users and satellites by evaluating whether the corresponding ray is blocked by environmental objects. The overall workflow, as illustrated in Fig. \ref{fig_2}, comprises three stages: 
\begin{itemize}
%\subsubsection{
\item {\em Data construction}: including satellite ephemeris generation and user location initialization.

%\subsubsection
\item {\em DT environment establishment}: where realistic 3D scenes are constructed using Blender with geographic data imported from OpenStreetMap. The resulting environment explicitly includes building geometries, which are integrated into Sionna’s ray-tracing engine.

%\subsubsection
\item {\em Visibility window computation}: where ray tracing is performed to determine LOS/NLOS conditions between each user–satellite pair over time. If the direct path is obstructed by terrain or buildings, the link is classified as NLOS; otherwise, it is considered visible.
\end{itemize}

Compared with geometry-based visibility models relying solely on elevation-angle thresholds, the ray-tracing-based DT explicitly captures blockage effects caused by realistic obstacles. This enables more accurate visibility determination in realistic environments, thereby providing a more faithful representation of user–satellite LOS dynamics. By leveraging orbital parameters and environmental contexts, the gateway performs forward-looking predictions to update user-satellite visibility windows at a predefined periodic interval $T_s$. The mechanism effectively mitigates the cumulative drift arising from LEO and user dynamics, thereby ensuring high-precision real-time scheduling while maintaining signaling efficiency.

\subsection{Traffic Model}
For the $u$-th ground user, the corresponding service queue $\mathbb{Q}_{u}^{t}$ is stored on the LEO satellite. Specifically, each service queue comprises multiple types of traffic, denoted as $\mathbb{Q}_{u}^{t}=\{\mathbb{Q}_{R{{T}_{1}},u}^{t}\ldots \mathbb{Q}_{R{{T}_{C}},u}^{t},\mathbb{Q}_{NRT,u}^{t}\}$, where $C$ is the number of RT service types.

For the $c$-th class RT service queue $\mathbb{Q}_{R{{T}_{c}},u}^{t}$ of user $u$ at time slot $t$, the arrival traffic $\Phi _{R{{T}_{c}},u}^{t}$ follows a Poisson distribution with arrival rate $\lambda _{R{{T}_{c}},u}^{t}$. ${{T}_{ttl,R{{T}_{c}}}}$ is defined as the Time to Live (TTL), which means that a packet will be discarded if it remains in the queue $\mathbb{Q}_{R{{T}_{c}},u}^{t}$ for more than ${{T}_{ttl,R{{T}_{c}}}}$. Accordingly, the amount of the $c$-th class RT service in $\mathbb{Q}_{R{{T}_{c}},u}^{t}$ is calculated as $d_{R{{T}_{c}},u}^{t}=\sum\nolimits_{l=t-{{T}_{ttl,R{{T}_{c}}}}+1}^{t}{\Phi _{R{{T}_{c}},u}^{l}}$,  and the total traffic for user $u$ in the satellite queue is then obtained as $d_{u}^{t}=\sum\limits_{c=1}^{C}{d_{R{{T}_{c}},u}^{t}}+d_{NRT,u}^{t}$.

\subsection{Communication Model}
It is assumed that each beam is consistently steered toward the center of its served beam position, and a widely adopted channel model is used for the BH LEO satellite downlink \cite{35,36}. The link gain between the $k$-th beam and the $u$-th user at time slot $t$ can be expressed as:
\begin{equation}
\label{eq1}
{{\left| h_{ku}^{t} \right|}^{2}}=\frac{G_{tx,ku}^{t}G_{rx,ku}^{t}}{{{(4\pi d_{u}^{t}f/c)}^{2}}},k\in \mathcal{K}^t,u\in \mathcal{U},
\end{equation}
where $d_u^t$ is the propagation distance from the satellite to the 
$u$-th user, $c$ is the speed of light, and $f$ is the downlink frequency.  The transmit antenna gain of the beam serving the $k$-th beam position for the $u$-th user at time slot $t$ is denoted by $G_{tx,ku}^t$, while $G_{rx,ku}^t$ represents the corresponding receive antenna gain. The gains are determined by the radiation patterns of the satellite and the user terminal, respectively. Given that the user’s receiving antenna continuously tracks the satellite direction, we 
%obtain 
have 
$G_{rx,ku}^t=G_{rx,max}$.

Considering the co-channel interference between intra-satellite beams, the signal-to-interference-noise ratio (SINR) of the $u$-th user served by the $k$-th beam in time slot $t$ can be formulated as:
\begin{equation}
\label{eq2}
SINR_{ku}^{t}=\frac{{{\left| h_{ku}^{t} \right|}^{2}}P_{ku}^{t}}{\sigma {{_{ku}^{t}}^{2}}+\sum\limits_{{k}^{'}=1,{{k}^{'}}\ne k}^{K}{{{\left| h_{{{k}^{'}}u}^{t} \right|}^{2}}P_{{{k}^{'}}u}^{t}}},
\end{equation}
where $\sigma {{_{ku}^{t}}^{2}}=\kappa {{T}_{noise}}B_{ku}^{t}$, $\kappa$ is the Boltzmann constant, and ${T}_{noise}$ is the noise temperature. Furthermore, $B_{ku}^{t}=N_{sub,ku}^{t}B/{{N}_{sub}}$, where ${N}_{sub}$ is the number of allocable subchannels within each beam, $N_{sub,ku}^{t}$ is the number of subchannels allocated to 
the $u$-th user in the $k$-th beam in time slot $t$, and $B$ is the beam bandwidth. With beam power evenly distributed among subchannels, we have $P_{ku}^{t}=P_{k}^{t}/{{N}_{sub}}*N_{sub,ku}^{t}$.

On this basis, according to Shannon's theory, the achievable downlink transmission rate of the $u$-th user is
\begin{equation}
\label{eq3}
c_{u}^{t}=\sum\limits_{k=1}^{K}{\mathbb{I}(y_{ku}^{t}=1)B_{ku}^{t}{{\log }_{2}}(1+SINR_{ku}^{t})}.
\end{equation}

In Eq. \ref{eq3}, 
$\mathbb{I}(\bullet )$ denotes the indicator function, where $\mathbb{I}(y_{ku}^{t}=1)=1$ if the $u$-th user is served by the $k$-th beam in time slot $t$, and $\mathbb{I}(y_{ku}^{t}=1)=0$ otherwise. Considering the service queue of the corresponding user on the satellite, the transmission throughput is $Th_{u}^{t}=\min \{d_{u}^{t},c_{u}^{t}\Delta t\}$.
\subsection{Problem Formulation}
In a multi-user BH LEO satellite system supporting differentiated services, the limited onboard resources necessitate enhancing energy efficiency to reduce power consumption and prolong the satellite’s operational lifetime. To satisfy the QoS requirements of diverse services, the system seeks to maximize the transmission of RT services while simultaneously ensuring fairness among users as a separate objective. These considerations motivate the formulation of a multi-objective optimization problem.
\begin{equation}
\label{eq4}
\begin{aligned}
   opt.{{\mathcal{P}}_{1}}&=\max \sum\limits_{t\in \mathcal{T}}{\sum\limits_{u\in \mathcal{U}}{Th_{u}^{t}}}/\sum\limits_{k=1}^{K}{\sum\limits_{u\in \mathcal{U}}{y_{ku}^{t}P_{ku}^{t}}} \\ 
       {{\mathcal{P}}_{2}}&=\max \sum\limits_{t\in \mathcal{T}}{\sum\limits_{u\in \mathcal{U}}{Th_{RT,u}^{t}}} \\ 
       {{\mathcal{P}}_{3}}&=\min \sum\limits_{t\in \mathcal{T}}{std{{\{\frac{\bar{R}_{u}^{t}}{\hat{R}_{u}^{t}}\}}_{u\in \mathcal{U}}}} \\ 
       \mathcal{P}&={{\omega }_{1}}{{\mathcal{P}}_{1}}+{{\omega }_{2}}{{\mathcal{P}}_{2}}+{{\omega }_{3}}{{\mathcal{P}}_{3}}, {{\omega }_{o}}\in [0,1], and \\
       &\sum\limits_{o=1}^{3}{{{\omega }_{o}}}=1, \\ 
            s.t. &C1:\sum\limits_{b\in \mathcal{B}}{x_{b}^{t}\le K,} x_{b}^{t}\in \{0,1\}, \\ 
 &               C2:\sum\limits_{k=1}^{K}{\sum\limits_{u\in \mathcal{U}}{y_{ku}^{t}P_{ku}^{t}\le {{P}_{\max }}, }}y_{ku}^{t}\in \{0,1\},P_{ku}^{t}\ge 0, \\ 
 &               C3:\sum\limits_{k=1}^{K}{\sum\limits_{u\in \mathcal{U}}{y_{ku}^{t}P_{ku}^{t}\ge \eta {{P}_{\max }}, }} \\ 
 &               C4:\sum\limits_{u\in \mathcal{U}}{N_{sub,ku}^{t}\le {{N}_{sub}},k \in \{1,2\ldots ,K\}.} \\ 
\end{aligned}
\end{equation}

In 
%Problem (4), 
Eq. \ref{eq4}, 
${{\mathcal{P}}_{1}}$ is formulated to optimize the energy efficiency of the LEO satellite, where $P_{max}$ denotes the total available onboard power. ${{\mathcal{P}}_{2}}$ aims to maximize the system’s RT service throughput, while ${{\mathcal{P}}_{3}}$ is designed to enhance user service fairness by minimizing the standard deviation of user satisfaction. The associated constraints are as follows: $C1$ requires the number of beams serving users 
%to 
be 
at most $K$; $C2$ ensures that the total power does not exceed the satellite’s maximum available power. To balance both energy and spectral efficiency, $C3$ enforces a minimum total power allocation $\eta {{P}_{\max }}$ for the serving beams, where $\eta$ denotes the power control factor; and $C4$ restricts the number of subchannels allocated to users within each beam to the total number of available system subchannels.

\section{Joint Beam Scheduling and Power Optimization for BH LEO Satellite Systems with Differentiated Users}
\label{sec4}
Directly solving the mixed-integer joint optimization in Eq. \ref{eq4} is hindered by the combinatorial explosion of the action space and the stringent real-time requirements of LEO satellites. To circumvent this intractability, we propose a hierarchical decomposition that separates the problem into beam-level and intra-beam scheduling stages. Specifically, the intra-beam subchannel assignment is addressed through a computationally lean, QoS-driven greedy method \cite{46} defined by the following priority metric: 
\begin{equation}
\label{eq17}
p_{tra,b,{{u}_{b}}}^{t}=\left\{ \begin{aligned}
  & -\frac{\log {{\delta }_{tra}}}{{{\tau }_{tra}}}\frac{d_{tra,b,{{u}_{b}}}^{t}HoL_{tra,b,{{u}_{b}}}^{t}Th_{tra,b,{{u}_{b}}}^{t}}{\hat{R}_{tra,b,{{u}_{b}}}^{t}}, \\ 
 &\ \ \ \ \ \ \ \ \  \  \ \ \                tra={{\{R{{T}_{c}}\}}_{c=1,...,C}}, \\ 
 & \frac{d_{tra,b,{{u}_{b}}}^{t}Th_{tra,b,{{u}_{b}}}^{t}}{\hat{R}_{tra,b,{{u}_{b}}}^{t}},tra=NRT, \\ 
\end{aligned} \right.
\end{equation}
where ${{\delta }_{tra}}$ represents the packet loss rate (PLR), $\tau_{tra}$ is the delay budget, and notably, $b\in {{\mathcal{K}}^{t}},{{u}_{b}}\in \{u|{{a}_{u,b=1}}\}$.

With the subchannel allocation strategy established, the remainder of this section focuses on the beam-level scheduling problem. The problem is firstly modeled as an MDP to capture the time-varying characteristics of the LEO satellite environment and the dynamic demands of users. We then introduce 
%the 
BH with Reinforcement learning incorporating Integrated Dirichlet and Gumbel-TopK Exploration (BRIDGE) algorithm. The framework effectively manages the complex hybrid discrete-continuous decision space and achieves significant improvements in system efficiency while ensuring the QoS requirements.

\subsection{MDP Formulation}
As described in 
%problem (
Eq. \ref{eq4}, beam scheduling and power allocation in BH LEO satellite systems can be regarded as a typical decision-making problem, where the resource allocation strategy adopted in each time slot determines the subsequent environment state. Consequently, the problem is modeled as an MDP. Nevertheless, due to the complexity of the environment and the challenge of accurately characterizing the state transition dynamics, a model-free DRL approach is employed, 
%A model-free MDP 
which consists of triples $(\mathcal{S},\mathcal{A},R)$, where $\mathcal{S}$, $\mathcal{A}$ and $R$ denote the state space, action space, and reward function, respectively. At time slot $t$, the agent observes the environment state $s^t \in \mathcal{S}$ and selects an action $a^t$ according to its policy $\pi$. The environment then transitions to a new state $s^{t+1}$, and the agent receives a reward $R(s^t,a^t)$. The state space, action space, and reward function are defined as follows:
\subsubsection{State Space $\mathcal{S}$}
 The observed state $s^t$ in each time slot is the state set of each user in the onboard queue, and $s_u^t$ is defined as:
\begin{equation}
\label{eq5}
s_{u}^{t}=\{Q_{u}^{t},H_{u}^{t},L_{u}^{t},I_{u}^{t},W_{u}^{t}\}.
\end{equation}

Specifically,
\begin{itemize}
\item $Q_{u}^{t}$: $6(C+1)$-dimension queue state, including the amount of data, latency, data arrival rate, average throughput, as well as the head-of-line (HoL) status of RT services and the traffic demand in the initial time slot.
\item $H_{u}^{t}$: Link gain of the $u$-th user.
\item $L_{u}^{t}$: The latitude and longitude coordinates of the $u$-th user’s beam position relative to the LEO sub-satellite point, denoted by $L_{u}^{t}=(Lat_{u}^{t},Lon_{u}^{t})$.
\item$I_{u}^{t}$: Beam index of the $u$-th user.
\item$W_{u}^{t}$: The remaining service time of the LEO satellite.
\end{itemize}

\subsubsection{Action Space $\mathcal{A}$} Based on the observed state, the agent performs a joint action of beam scheduling and power allocation
\begin{equation}
\label{eq6}
{{a}^{t}}=\{{{X}^{t}},{{P}^{t}}\},
\end{equation}
where beam scheduling ${{X}^{t}}={{\{x_{b}^{t}\}}_{b\in \mathcal{B}}}$ is defined over a discrete action space, 
%while 
and power allocation ${{P}^{t}}={{\{P_{k}^{t}\}}_{k=1,...,K}}$ is defined over a continuous action space.

\subsubsection{ Reward Function $R$} Considering the objective function of the optimization problem in (Eq. \ref{eq4}), a linear weighted multi-objective modeling approach is adopted to design the reward function for DRL training. Specifically, a reward function ${{r}^{t}}={{\omega }_{1}}r_{1}^{t}+{{\omega }_{2}}r_{2}^{t}+{{\omega }_{3}}r_{3}^{t}$ is given by
\begin{equation}
\label{eq7}
\begin{aligned}
  & r_{1}^{t}=\sum\limits_{u\in \mathcal{U}}{Th_{u}^{t}}/\sum\limits_{k=1}^{K}{\sum\limits_{u\in \mathcal{U}}{y_{ku}^{t}P_{ku}^{t}}} \\ 
 & r_{2}^{t}=\sum\limits_{u\in \mathcal{U}}{Th_{RT,u}^{t}} \\ 
 & r_{3}^{t}=-std{{\{\frac{\bar{R}_{u}^{t}}{\hat{R}_{u}^{t}}\}}_{u\in \mathcal{U}}} \\ 
\end{aligned}.
\end{equation}

Notably, since $\mathcal{P}_3$ represents a minimization problem, a negative sign is applied to its corresponding sub-reward function. The linear weighted formulation enables controllable trade-offs among multiple objectives by adjusting the weighting factors, thereby transforming the original multi-objective optimization problem into a standard single-objective problem, which is widely adopted in DRL-based optimization \cite{41}.

\subsection{Algorithm Design}
To address the hybrid discrete-continuous action space, we develop BRIDGE, a DRL framework built upon PPO. The selection of beam subsets leads to an exponential expansion of the discrete action space as the number of beam positions increases, posing significant challenges to the search efficiency and convergence of conventional DRL schemes. To mitigate the curse of dimensionality, BRIDGE incorporates a Gumbel-TopK sampling mechanism for discrete beam decision-making, enabling efficient and differentiable exploration within a vast combinatorial space. Concurrently, the Dirichlet distribution is employed to model continuous power allocation actions, which internalizes non-negativity and total power budget constraints through probabilistic mapping. This architecture significantly enhances the capability for resource allocation in extreme high-dimensional scenarios while ensuring learning stability and convergence under complex constraints. The overall network architecture and training procedure are summarized below.

\begin{figure*}[!t]
\centering
\includegraphics[width=0.9\textwidth]{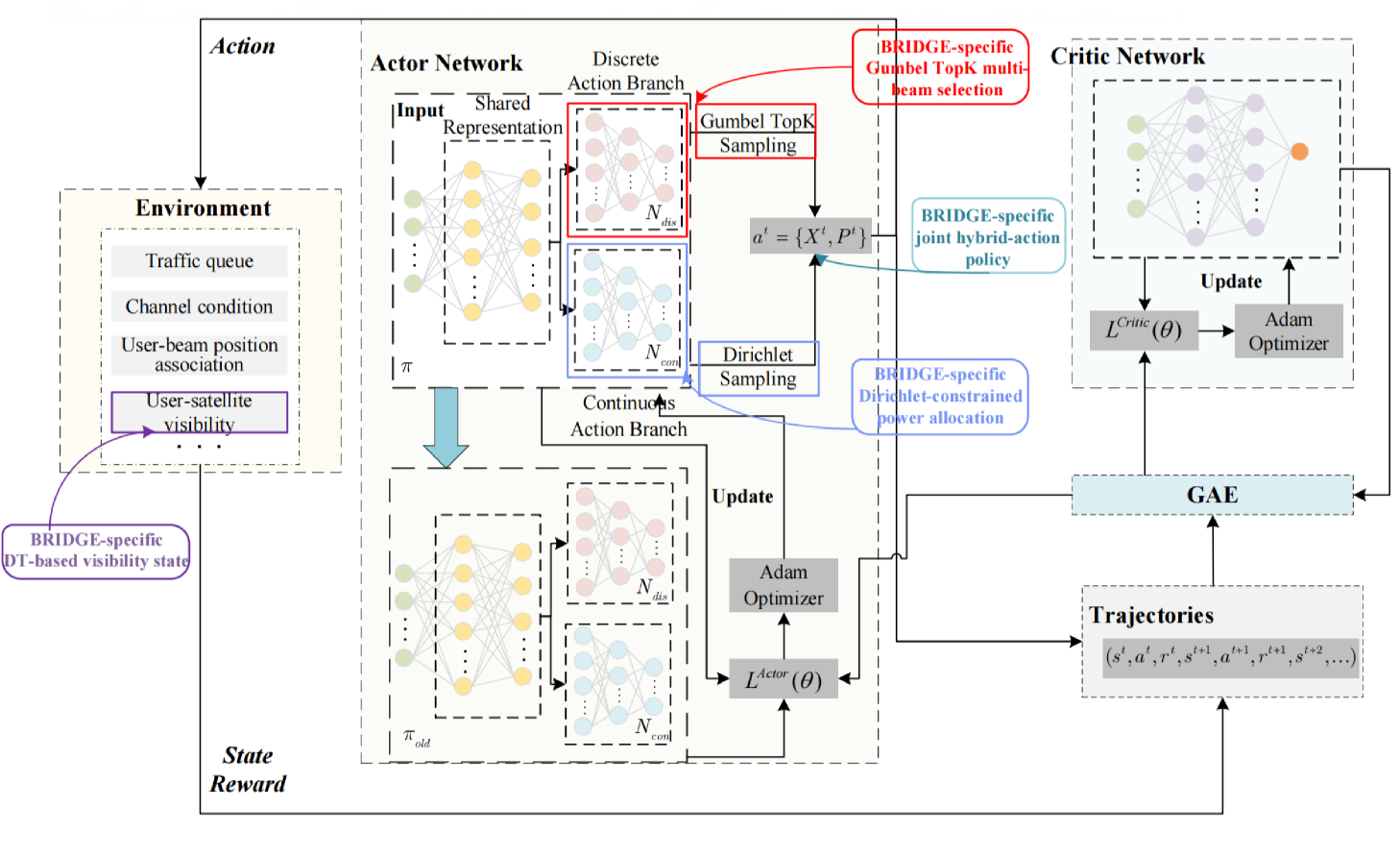}
\vspace{-4mm}
\caption{Network Architecture of the proposed BRIDGE algorithm.}
\label{fig_3}
\end{figure*}

\subsubsection{Network Architecture}
The proposed network architecture is shown in Fig. \ref{fig_3}, which adopts the actor-critic PPO framework. The actor network receives the input state, which first passes through a shared representation layer—a multi-layer fully connected network designed to extract common features from the high-dimensional state. The resulting shared encoding is then forwarded to two separate branches:
\begin{itemize}
\item \textit{Discrete Action Branch}: This branch is responsible for beam scheduling. It generates a candidate action distribution via several fully connected layers and employs Gumbel-TopK sampling to enable structured exploration over the large combinatorial beam selection space, allowing multiple beams to be activated simultaneously at each decision step. The independent Gumbel noise $g_i$ is added to the outputs of the $N_{dis}$-dimensional discrete action branch network, specifically:
\begin{equation}
\label{eq8}
{{g}_{i}}=-\log (-\log ({{n}_{i}})),{{n}_{i}}\sim Uniform(0,1).
\end{equation}

The TopK largest values are then selected, and their corresponding indices are used as the served beam positions:
\begin{equation}
\label{eq9}
\mathcal{B}_{s}^{t}=\arg top{{k}_{i=\{1,...,{{N}_{dis}}\}}}({{o}_{i,dis}}+{{g}_{i}}).
\end{equation}

\item \textit{Continuous Action Branch}: The branch is responsible for power allocation. Conditioned on the selected beam set, the Dirichlet-sampled power allocation dynamically distributes the available onboard power across the active beams at each time step, ensuring feasibility while enabling continuous control. This branch produces an $N_{con}=K+1$-dimensional parameterized vector via fully connected layers, from which continuous actions $P_D^t$ are sampled using the Dirichlet distribution to inherently satisfy non-negativity and the total power constraint. In this work, power allocation is further refined based on an equal-power baseline, thereby
\begin{equation}
\label{eq10}
{{P}^{t}}={{P}_{\max }}(\eta /|{{\mathcal{K}}^{t}}|+(1-\eta )P_D^t).
\end{equation}

\end{itemize}

Additionally, the critic network is employed to learn the state-value function. After processing through the hidden and output layers, it directly produces the corresponding state value $V(s^t)$.

\subsubsection{Training Procedure}
The proposed algorithm observes state $s^t$ at time slot $t$ and interacts with the environment based on the actor network's policy $\pi_{\theta}$. It obtains the reward $r^t$ and the next state ${{s}^{t+1}}$ and stores them to form a trajectory data $({{s}^{t}},{{a}^{t}},{{r}^{t}},{{s}^{t+1}})$. The policy ${{\pi }_{\theta }}({{a}^{t}}|{{s}^{t}})$ can be expressed as:
\begin{equation}
\label{eq11}
{{\pi }_{\theta }}({{a}^{t}}|{{s}^{t}})={{\pi }_{{{\theta }_{dis}}}}(\mathcal{B}_{s}^{t}|{{s}^{t}})*{{\pi }_{{{\theta }_{con}}}}({{P}^{t}}|{{s}^{t}}).
\end{equation}

Furthermore, as described in the network structure, ${{\pi }_{{{\theta }_{con}}}}({{P}^{t}}|{{s}^{t}})$ is obtained by sampling the Dirichlet distribution \cite{37}, whereas ${{\pi }_{{{\theta }_{dis}}}}(\mathcal{B}_{s}^{t}|{{s}^{t}})$ is generated by sampling the Gumbel-TopK distribution, with its probability calculated according to the Plackett–Luce distribution \cite{38}. Specifically,
\begin{equation}
\label{eq12}
{{\pi }_{{{\theta }_{dis}}}}(\mathcal{B}_{s}^{t}|{{s}^{t}})=\prod\limits_{i=1}^{k}{\frac{exp({{o}_{{{\sigma }_{i}},dis}})}{\sum\nolimits_{j=i}^{{{N}_{dis}}}{exp({{o}_{{{\sigma }_{j}},dis}}})}},
\end{equation}
where $\sigma =({{\sigma }_{1}},{{\sigma }_{2}},\ldots ,{{\sigma }_{{{N}_{dis}}}})$ denotes the sorting indices of the discrete outputs of the action network after adding noise. To ensure that the actor network updates not only maximize the expected return but also maintain training stability, a clipped objective function is employed to prevent excessive gradient updates:
\begin{equation}
\label{eq13}
\begin{aligned}
  & {{L}^{clip}}(\theta )= \\ 
 & {{\mathbb{E}}_{t}}[\min (rati{{o}_{t}}(\theta ){{A}_{t}},clip(rati{{o}_{t}}(\theta ),1-\epsilon ,1+\epsilon ){{A}_{t}})]. \\ 
\end{aligned}
\end{equation}

In the above formula, $rati{{o}_{t}}(\theta )={{\pi }_{\theta }}({{a}^{t}}|{{s}^{t}})/{{\pi }_{{{\theta }_{old}}}}({{a}^{t}}|{{s}^{t}})$, and the advantage function $A_t$ is calculated by the generalized advantage estimation (GAE) function:
\begin{equation}
\label{eq14}
{{A}_{t}}=\sum\limits_{l=0}^{T-1-t}{{{(\gamma \lambda )}^{l}}{{\delta }_{t+l}}},
\end{equation}
where $\gamma$ denotes the discount factor, and $\lambda$ is the GAE smoothing coefficient, which balances the bias and variance of the advantage estimation, with ${{\delta }_{t}}={{r}^{t}}+\gamma V({{s}^{t+1}})-V({{s}^{t}})$.  To enhance policy exploration and prevent premature convergence, an entropy regularization term is also included in the loss function
\begin{equation}
\label{eq15}
{{L}^{Actor}}(\theta )=-({{L}^{clip}}(\theta )+c_e{{\mathbb{E}}_{t}}[\mathcal{H}({{\pi }_{\theta }}(\cdot |{{s}^{t}}))].
\end{equation}

The update goal of the critic network is to minimize the mean square error between the predicted value function and the actual return, that is,
\begin{equation}
\label{eq16}
{{L}^{Critic}}(\phi )={{\mathbb{E}}_{t}}{{[{{V}_{\phi }}({{s}^{t}})-({{V}_{{{\phi }_{old}}}}({{s}^{t}})+{{A}_{t}})]}^{2}}.
\end{equation}

Algorithm 
%1 
\ref{alg} 
presents the pseudocode of the proposed BRIDGE algorithm. Initially, the parameters $\theta$ and $\phi$ of the actor and critic networks are initialized. At each time step $t$ within an episode, given the current state $s^t$, the agent produces an action distribution by the actor network. The beam scheduling vector $\mathcal{B}_{s}^{t}$ is sampled from the discrete action branch, while the beam power vector $P^t$ is obtained from the continuous action branch. Based on the selected beam schedule, the served beam positions $\mathcal{K}^t$ at time slot $t$ are determined. For each beam, the scheduling priority of individual user services $p_{tra,b,{{u}{b}}}^{t}$ is computed, services are greedily allocated, and the remaining subchannels $N_{sub,res}$ are updated until all beams are processed. After each time step, the immediate reward $r^t$ is obtained, and the environment transitions to the next state $s^{t+1}$. Upon episode completion, the critic network estimates the value of each state $V(s^t)$, and the advantage function $A^t$ at each time step is computed using the discounted reward combined with GAE. Finally, the calculated advantage function is used to update the actor network parameters, achieving joint optimization of both actor and critic networks.

Dirichlet sampling and Gumbel-TopK exploration are applied only at the action sampling stage and do not modify the underlying policy gradient update mechanism, thereby preserving the theoretical structure and convergence of PPO. In addition, the proposed BRIDGE algorithm is trained offline on ground-based servers with high computational capability. After training, the learned network parameters are updated to the satellite via the feeder link. During operation, the trained resource management model performs online inference to support resource allocation decisions for LEO satellites, which is considered feasible for practical deployment.
\begin{algorithm}
\caption{BH with Reinforcement learning incorporating Integrated Dirichlet and Gumbel-TopK Exploration (BRIDGE) Algorithm.}
\begin{algorithmic}[1]
\STATE Initialize actor and critic networks with random parameters $\theta = \theta^{-}$ and $\phi = \phi^{-}$.
\STATE Initialize the trajectory buffer and environment state.
\FOR{each episode}
    \STATE Reset environment and obtain initial state $s^0$.
    \FOR{each time step $t$}
        \STATE Observe current state $s^t$, and select $\mathcal{B}_{s}^{t}\sim {{\pi }_{{{\theta }_{dis}}}}({{s}^{t}})$ and ${{P}^{t}}\sim {{\pi }_{{{\theta }_{con}}}}({{s}^{t}})$.
        \STATE Determine served beam positions $\mathcal{K}^t$ based on $\mathcal{B}_{s}^{t}$.
        \FOR{each beam $do$}
            \STATE  ${{N}_{sub,res}}\leftarrow{{N}_{sub}}$.
            \WHILE{${{N}_{sub,res}}>0$ and $\sum\nolimits_{{{u}_{b}}\in \{u|{{c}_{u,b=1}}\}}{d_{u}^{t}}>0$}
            \STATE Establish scheduling priority of each user served $p_{tra,b,{{u}_{b}}}^{t}$ according to Eq.(\ref{eq17}).
            \STATE $tr{{a}_{0}},{{u}_{b,0}}\leftarrow \underset{tra,{{u}_{b}}}{\mathop{\arg \max }}\,{{p}_{tra,b,{{u}_{b}}}}$.
            \STATE${{N}_{sub,res}}\leftarrow {{N}_{sub,res}}-{{N}_{sub,k{{u}_{b}}}}$.
            \ENDWHILE
        \ENDFOR
        \STATE Obtain immediate reward $r^t$.
        \STATE Update environment state $s^{t+1}$.
    \ENDFOR
    \STATE Estimate state value $V(s^t)$ using critic network.
    \STATE Compute advantage function $A^t  $ using GAE.
    \STATE Update actor network parameter $\theta$ and critic network $\phi$.
\ENDFOR
\end{algorithmic}
\label{alg}
\end{algorithm}
\subsection{Complexity Analysis}
%The subsection focuses 
We focus on the inference phase of the proposed algorithm, as this stage determines its real-time performance during deployment. The dominant cost arises from the forward propagation through the actor network. The network architecture can be divided into three computational components: a shared feature extraction module with $L_s$ layers, a beam selection branch containing $L_b$ layers, and a power allocation branch with $L_p$ layers. The overall inference complexity is expressed as:
\begin{equation}
\label{eq18}
\begin{aligned}
  & \mathcal{O}({{d}_{i}}{{d}_{s1}}+\sum\limits_{i=1}^{{{L}_{s}}-1}{{{d}_{si}}{{d}_{s(i+1)}}}+{{d}_{s{{L}_{s}}}}({{d}_{b1}}+{{d}_{p1}}) \\ 
 & +\sum\limits_{j=1}^{{{L}_{b}}-1}{{{d}_{bj}}{{d}_{b(j+1)}}}+\sum\limits_{k=1}^{{{L}_{p}}-1}{{{d}_{pk}}{{d}_{p(k+1)}}}) \\ 
\end{aligned},
\end{equation}
where $d_i$ denotes the input layer dimension, while $d_{si}$, $d_{bj}$, and $d_{pk}$ correspond to the layer dimensions of the shared feature extractor, beam selection branch, and power allocation branch, respectively. Additional operations, such as Gumbel-TopK sampling and Dirichlet sampling, incur negligible complexity compared to the dominant matrix multiplications. 

The computational complexity of the subchannel allocation algorithm for differentiated QoS requirements, where $U_b$ denotes the maximum number of users per beam, can be approximated as:
\begin{equation}
\label{eq19}
\mathcal{O}(K{{U}_{b}}\min \{(C+1){{U}_{b}},{{N}_{sub}}\}).
\end{equation}

Compared to the DRL inference complexity in  Eq. \ref{eq18}, the computational cost of this step remains lower under typical system configurations with a moderate number of users per beam. However, when the user density per beam becomes large, this cost may no longer be negligible.

\section{Adversarial Attacks Against BRIDGE-Based Beam Scheduling and Power Allocation in LEO Satellite Systems}
\label{sec5}

This section presents the adversarial attack models and algorithms considered for evaluating the vulnerability of the proposed BRIDGE algorithm. Specifically, widely used gradient-based attack methods, including FGSM, I-FGSM, and PGD, are applied to quantitatively assess the robustness of the model under adversarial environments. The resulting analysis provides a foundation for performance evaluation and offers insights for enhancing the resilience of intelligent scheduling in LEO satellite systems. 

\subsection{Attack Model}

\begin{figure}[!t]
\centering
\includegraphics[width=0.9\columnwidth]{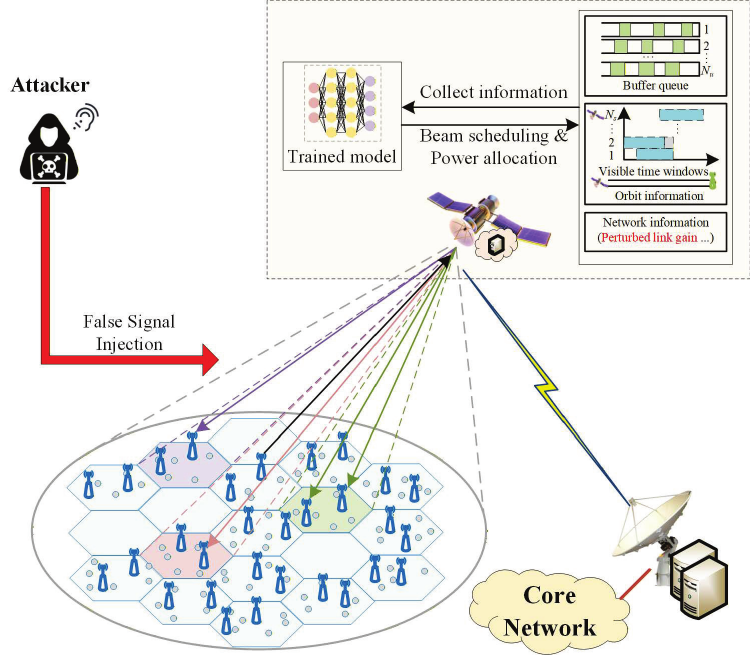}
\vspace{-4mm}
\caption{Illustration of adversarial attack against the proposed AI-enabled resource allocation model.}
\label{fig_4}
\end{figure}

The BRIDGE algorithm dynamically allocates satellite resources within a hybrid action space according to link conditions and traffic demands, ensuring QoS requirements for heterogeneous services under normal operation. To evaluate robustness under abnormal or malicious conditions, we design an adversarial attack scenario targeting the scheduling model. As illustrated in Fig. \ref{fig_4}, the attacker perturbs the perceived link gain state at the satellite, inducing the agent to make suboptimal or unstable resource allocation decisions and thereby revealing potential vulnerabilities under perturbed inputs. It should be clarified that the assumed perturbation of link gain does not imply any direct physical manipulation of the satellite payload or the propagation channel. Rather, the adopted attack model serves as an abstraction of potential corruption or uncertainty in the channel state information (CSI) available at the satellite. Such distortions may arise from falsified uplink CSI feedback, misleading or noisy measurements, estimation errors at the gateway or network control plane, as well as inaccuracies in Digital Twin–based channel or visibility prediction. From the perspective of the scheduling agent, these factors manifest as perturbations in the perceived link gain state and can therefore be equivalently modeled as input disturbances.

To maintain both stealthiness and analytical tractability, the perturbations are strictly constrained within acceptable bounds. While practical deployment of such attacks may be challenging, their theoretical formulation provides a rigorous framework for assessing the robustness of learning-based scheduling models. This analysis allows evaluation of performance degradation under adversarial conditions and offers guidance for designing defense strategies to enhance the security and reliability of BH LEO satellite resource allocation systems.

\subsection{Adversarial Attack Methods}
Considering BH LEO satellite resource allocation, adversarial attacks involve introducing carefully crafted, small perturbations into the observed link states to systematically bias the BRIDGE algorithm’s decisions, primarily because such imperceptible deviations bypass standard threshold-based detectors that would otherwise intercept large-magnitude statistical outliers. Such perturbations can lead to suboptimal beam scheduling and power allocation, thereby degrading overall system performance. While conventional attacks typically use additive perturbations, $x+{{\epsilon }^{adv}}$, multiplicative perturbations, $x(1+{{\epsilon }^{adv}})$, scale the input proportionally, allowing the perturbation strength to adapt to the input amplitude and improving both attack effectiveness and stability. 
While more sophisticated adaptive attack strategies are of interest, their modeling typically requires additional assumptions on attacker observability and interaction dynamics, which are beyond the scope of the work. Common adversarial attack algorithms applied in this context include FGSM, I-FGSM, and PGD, each employing specific strategies for generating perturbations. These methods are selected to evaluate fixed-budget robustness under bounded link gain perturbations. Together, they provide a progressive evaluation from single-step attacks to stronger iterative attacks under the same bounded perturbation model, which is different from minimal-distortion attacks such as C\&W-style attacks.

Table \ref{tab_1} 
summarizes the differences among these methods, with detailed implementations provided below.

\begin{table}[htbp]
\centering
\caption{Comparison of FGSM, I-FGSM, and PGD Methods}
\begin{tabular}{|c|c|c|c|}

\hline
\textbf{Method} & \textbf{Initial Point} & \textbf{Iterative} & \textbf{Attack Strength} \\
\hline
FGSM & Original Point & No & Weak \\
\hline
I-FGSM & Original Point & Yes & Medium \\
\hline
PGD & Random Point  & Yes & Strong \\
\hline
\end{tabular}
\label{tab_1}
\end{table}

\subsubsection{FGSM Method} 
%The 
FGSM is a fundamental technique in adversarial attack research, enabling the efficient generation of adversarial examples. It calculates the gradient of the model’s loss function with respect to the input and applies a single-step perturbation in the direction that maximally increases the loss, thereby causing the model to produce incorrect predictions. 
Accordingly, the link gain states $\mathcal{H}_u^t$ observed by the satellite are perturbed as follows:
\begin{equation}
\label{eq20}
H_{u}^{adv,t}=H_{u}^{t}*(1+{{\epsilon }^{adv}}\cdot sign({{\nabla }_{H_{u}^{t}}}J(\theta ,s{}^{t},{{a}^{t}}))),
\end{equation}
where ${\epsilon }^{adv}$ is a hyperparameter that controls the magnitude of the perturbation, constraining the deviation between the adversarial sample and the original input. The symbol $sign(\cdot)$ denotes the sign function, which determines the direction of the perturbation. Accordingly, the adversarial example is generated by maximizing the objective $J$ as follows:
\begin{equation}
\label{eq21}
J(\theta ,{{s}^{t}},{{a}^{t}})=-(\log ({{\pi }_{{{\theta }_{dis}}}}(\mathcal{B}_{s}^{t}|{{s}^{t}}))+\log ({{\pi }_{{{\theta }_{con}}}}({{P}^{t}}|{{s}^{t}})).
\end{equation}

Intuitively, 
%the 
FGSM perturbation aligns with the direction of steepest ascent of the loss function. Thus, even perturbations of minimal magnitude can shift the input across the model’s decision boundary, resulting in bad policy outputs.

\subsubsection{I-FGSM Method} 
%The 
I-FGSM represents an extension of FGSM, designed to generate stronger adversarial examples by incrementally approaching the locally optimal adversarial direction through multiple small perturbation steps. Specifically, the adversarial sample is initialized as $H_{0,u}^{\text{adv},t}=H_{u}^{t}$, and at each iteration, it is updated by taking a small step ${{\alpha}^{\text{adv}}}$ along the direction of the gradient sign:
\begin{equation}
\label{eq22}
\begin{aligned}
  & H_{l+1,u}^{adv,t}= \\ 
 & cli{{p}_{H_{u}^{t},{{\epsilon }^{adv}}}}(H_{l,u}^{adv,t}(1+{{\alpha }^{adv}}sign({{\nabla }_{H_{l,u}^{adv,t}}}J(\theta ,s_{l}^{adv,t},{{a}^{t}})))). \\ 
\end{aligned}
\end{equation}

In Eq. \ref{eq22}, the clipping operation constrains the cumulative perturbation within the range $[1-{{\epsilon }^{\text{adv}}},1+{{\epsilon }^{\text{adv}}}]$. In contrast to FGSM, I-FGSM iteratively applies perturbations along the most sensitive gradient direction, enabling more precise alignment with the loss surface and thereby improving the attack success rate.
\subsubsection{PGD Method}
PGD builds on I-FGSM by introducing random initialization, enabling exploration of multiple starting points within the input space. The method increases the likelihood of crossing the model’s decision boundaries, generating more effective adversarial examples. The initial adversarial sample is created within a proportional neighborhood defined by ${{\epsilon }^{adv}}$ around the original input $H_u^t$:
\begin{equation}
\label{eq23}
H_{0,u}^{adv,t}=H_{u}^{t}(1+U(-{{\epsilon }^{adv}},{{\epsilon }^{adv}})),
\end{equation}
where $U(-{{\epsilon }^{adv}},{{\epsilon }^{adv}})$ denotes a value sampled uniformly from the interval $[-{\epsilon }^{adv},{\epsilon }^{adv}]$. Starting from this random initialization, the adversarial sample is iteratively updated along the gradient sign direction, with each step projected back into the permissible perturbation range. In contrast to I-FGSM, PGD employs multiple random initializations and selects the sample yielding the maximal loss after each iteration, thereby enhancing the attack’s effectiveness.

\section{Simulation Results and Performance Analysis}
\label{sec6}

\subsection{Simulation Parameters}

\begin{table}[!t]
\caption{Simulation Parameters}
\label{tab2}
\centering
\begin{tabular}{ll}%|c|l|
\hline
\textbf{Parameters}& \textbf{Values}\\
\hline
\textbf{Scenario parameters} & \\
%\hline
Altitude of the LEO satellite & 508 km\\
Ka band $f$ & 20 GHz\\
%\hline
Total bandwidth $B$ & 200 MHz\\
Number of subchannels $N_{sub}$ & 20\\
Number of users $N_u$ & 60\\
%\hline
Number of beams $K$&8\\
Noise temperature $T_{noise}$ & 290 K\\
%\hline
Power control factor $\eta$ & 0.8\\
Maximum transmit power ${{P}_{max}}$ & 250 W\\
%\hline
Maximum transmit antenna gain & 36.2 dBi\\
%\hline
3dB bandwidth of transmit antenna &$2.5{}^\circ $\\
%\hline
Maximum receive antenna gain & 20 dBi\\

Video service (RT) & 242 kbps with delay budget\\
& 300 ms and PLR budget $10^{-6}$\\
%\hline
Voice service (RT) & 8.4 kbps with delay budget\\
&100 ms and PLR budget $10^{-3}$\\
BE service (NRT) &100 kbps with delay budget\\
&1,000 ms \\
Time slot duration $\Delta t$&10 ms\\
%\hline
Satellite location update interval ${{T}_{s}}$ & 2 s\\
\textbf{Algorithm parameters}&\\
Learning rate of Actor Network & $3\times {{10}^{-5}}$\\
Learning rate of Critic Network & $3\times {{10}^{-5}}$\\
Discount factor $\gamma$ & 0.99\\
GAE smoothing coefficient $\lambda$&0.95 \\
Entropy regularization coefficient $c_e$&0.1\\
Clipping parameter $\epsilon$ & 0.2\\
\hline
\end{tabular}
\end{table}

In this work, a Ka-band BH LEO satellite communication system is simulated. Table \ref{tab2} summarizes the main simulation parameters, partially based on \cite{15,16}. The satellite orbital configuration is designed with reference to the GW satellite constellation, providing a realistic representation of large-scale LEO constellation. The system considers 60 ground users, with each satellite supporting up to 8 beams. A total bandwidth of 200 MHz is considered, equally divided into 20 subchannels. The power control factor $\eta$ is set to 0.8, and the maximum transmit power per satellite is limited to 250 W. Each user’s aggregate traffic demand follows a Poisson process, with the arrival rate determined by economic, population, and other factors\cite{44}. Three types of services are considered: two categories of RT services and one NRT service. The onboard queue is updated once per time slot, which lasts 10 ms. Maximum transmit and receive antenna gains are 36.2 and 20 dBi, respectively, and the orbital altitude of the LEO satellite is fixed at 508 km. To model individual beam positions within the coverage area, the open-source spatial indexing algorithm H3 developed by Uber is employed \cite{27}.

The training parameters are also listed in Table \ref{tab2}, with learning rates of $3\times 10^{-5}$ for both the actor and critic networks. Generalized Advantage Estimation (GAE) is applied using $\lambda=0.95$ and $\gamma=0.99$, while a clipping parameter $\epsilon=0.2$ and entropy regularization coefficient $c_e=0.1$ are adopted to stabilize policy updates and encourage exploration. The network architectures are detailed in Table \ref{tab3}. Specifically, the actor network comprises 782,321 parameters and involves a computational complexity of 1.56 MFLOPs per inference. While the theoretical raw computation time on a 275 TOPS platform \cite{45} is only approximately 5.67 ns, the empirical latency remains on the millisecond scale even after accounting for memory access and system overheads, thereby guaranteeing the real-time execution of beam management policies. Furthermore, the lightweight memory footprint of 2.98 MB is vital for resource-constrained LEO satellite payloads. A brief description of the comparison algorithms follows:

\begin{itemize}

%\subsubsection
\item {\em Queue Length Prioritized with Distance Limit Beam Hopping (QLPDL-BH)}:
Beam positions with the highest queue lengths are greedily scheduled, while ensuring that the distance between selected beams does not exceed the beam diameter \cite{11}.

%\subsubsection
\item {\em Periodic Beam Hopping (P-BH)}:
The P-BH method employs periodic BH, whereby all beam positions are served in a fixed, repeating sequence.

%\subsubsection
\item {\em Genetic Algorithm Beam Hopping (GA-BH)}:
Beam scheduling is optimized using a genetic algorithm with a population size of 50 individuals evolving over 50 generations \cite{39}.

%\subsubsection
\item {\em TopK DQN}:
The LEO satellite BH controller selects and serves the top $K$ beam positions based on their estimated Q-values \cite{17}.

%\subsubsection
\item {\em Soft Actor-Critic method-based Beam Hopping (SAC-BH)}:
At each time slot, the LEO satellite selects $K$ beam positions based on the SAC method to serve, with the total onboard power equally allocated among the active beams.

%\subsubsection
\item {\em Discrete PPO}: 
Beam scheduling decisions are determined by the proposed discrete action branch of the PPO-based BH controller.
\end{itemize}

\begin{table}[!htbp]
\centering
\caption{Network Architecture of Actor and Critic Network.}
\begin{tabular}{lcc}
\toprule
\textbf{Layer} & \textbf{Input Size} & \textbf{Output Size} \\
\midrule
\multicolumn{3}{l}{\textbf{Actor Network}} \\
Shared Representation Layer & & \\
\quad Layer1 & 1,200 & 512 \\
\quad Layer2 & 512  & 256 \\
Discrete Action Branch & & \\
\quad Layer1 & 256  & 64  \\
\quad Layer2 & 64   & 40  \\
Continuous Action Branch & & \\
\quad Layer1 & 256  & 64  \\
\quad Layer2 & 64   & 9   \\
\midrule
\multicolumn{3}{l}{\textbf{Critic Network}} \\
Layer1 & 1,200 & 512 \\
Layer2 & 512  & 128 \\
Layer3 & 128  & 1   \\
\bottomrule
\end{tabular}
\label{tab3}
\end{table}

\begin{figure*}[!t]
\centering
\includegraphics[width=\textwidth]{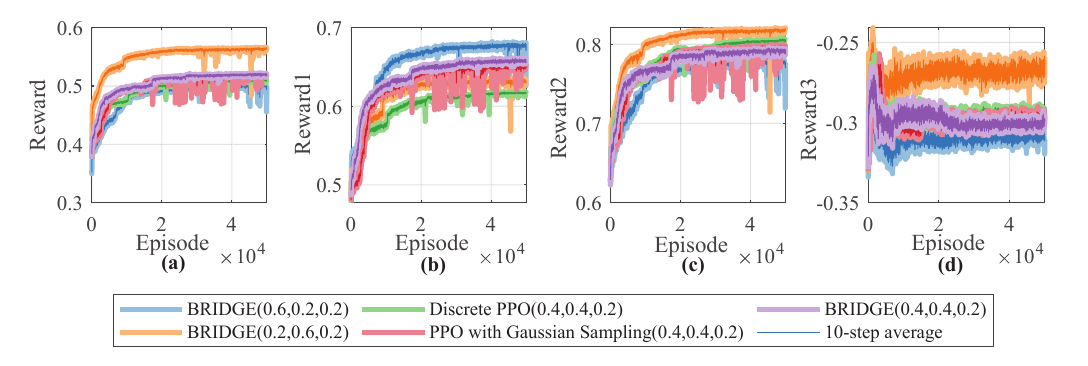}
\caption{Convergence analysis of the proposed BRIDGE algorithm: (a) total reward, (b) sub-reward 1, (c) sub-reward 2, and (d) sub-reward 3 under different optimization weight settings, compared 
%with 
to 
the discrete-only PPO and PPO with Gaussian sampling baseline.}
\label{fig_5}
\end{figure*}

\subsection{Convergence Analysis}
The proposed BRIDGE algorithm is trained offline. Fig. \ref{fig_5}(a)-(d) depict the convergence of the total reward and its three subcomponents under different optimization weight 
settings—(0.4,0.4,0.2), (0.6,0.2,0.2) and 
(0.2,0.6,0.2).  In addition, two baseline curves are included for comparison: a discrete-only optimization (denoted as “Discrete PPO”) and a PPO with Gaussian Sampling baseline, where continuous power allocation is modeled using Gaussian exploration while beam selection is handled by Gumbel-TopK trick. The total reward gradually increases and stabilizes across all weight configurations, demonstrating the reliable convergence of the proposed algorithm. The sub-reward curves converge consistently as well, indicating that the agents effectively balance multiple objectives. The “Discrete PPO” curve serves as a baseline, highlighting the performance gains achieved through joint beam scheduling and power allocation. While the PPO with Gaussian sampling baseline reaches a similar final reward level, it suffers from significantly larger convergence fluctuations, reflecting reduced stability compared with the proposed BRIDGE method. Overall, these results demonstrate that the BRIDGE algorithm achieves stable convergence while delivering superior performance under continuous–discrete cooperative optimization.

\subsection{Performance Analysis}
\subsubsection{Performance of Energy Efficiency}
\begin{figure*}[!t]
\centering
\includegraphics[width=0.8\textwidth]{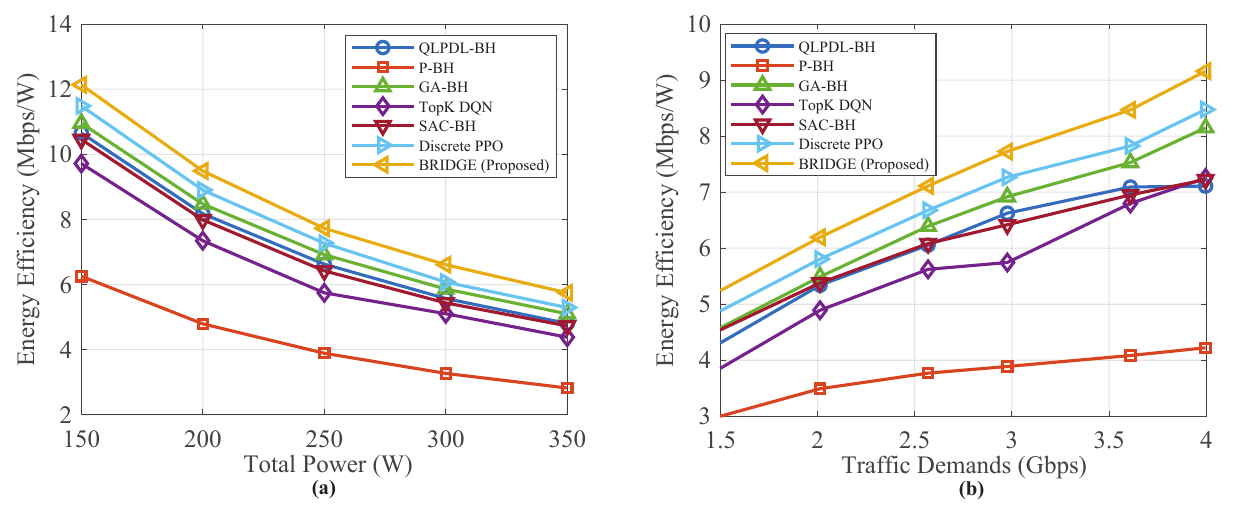}
\vspace{-4mm}
\caption{Energy Efficiency of different algorithms. (a) Energy Efficiency versus Total Power (Traffic Demand = 3 Gbps), (b) Energy efficiency versus Traffic Demands ($P_{max}$ = 250 W).}
\label{fig_6}
\end{figure*}

Fig. \ref{fig_6} illustrates the energy efficiency performance of different algorithms under varying on-board total power and traffic demand. The total on-board power ranges from 150 W to 350 W, and the overall traffic demand varies from 1.5 Gbps to 4 Gbps. For Fig. \ref{fig_6} - Fig. \ref{fig_9}, each algorithm (except P-BH and QLPDL-BH) was evaluated over 50 independent runs using a converged model, and the results reported correspond to the averaged performance across these runs. The results show that as the total power increases, the energy efficiency of all algorithms generally decreases. This is because higher power input does not yield proportional performance gains, leading to reduced throughput efficiency per unit of power consumption. In contrast, when traffic demand increases, overall energy efficiency improves, as higher traffic loads can more fully exploit the available power resources and enhance energy utilization.

Among the algorithms compared, BRIDGE and discrete PPO consistently rank among the top two in terms of energy efficiency. In particular, BRIDGE benefits from its flexible power allocation and scheduling mechanism, enabling adaptive optimization under varying traffic demands and power constraints. Specifically, compared with QLPDL-BH, P-BH, GA-BH, Top-K DQN, SAC-BH and discrete PPO, the proposed method improves energy efficiency by 16.62\%, 98.68\%, 11.68\%, 34.5\%, 15.41\%, and 6.32\%, respectively. Although Top-K DQN and SAC-BH can adapt to dynamic changes due to its learning-based framework, they are prone to being trapped in local optima and therefore struggles to achieve satisfactory convergence. GA-BH, as a heuristic algorithm with global optimization capability, outperforms other baselines, but its solutions require extensive search and iterative computations, limiting its real-time applicability. QLPDL-BH, by considering distance isolation among service beams, can mitigate interference to some extent and thus performs better than P-BH.

\subsubsection{Performance of RT Service Throughput}

\begin{figure*}[!t]
\centering
\includegraphics[width=0.8\textwidth]{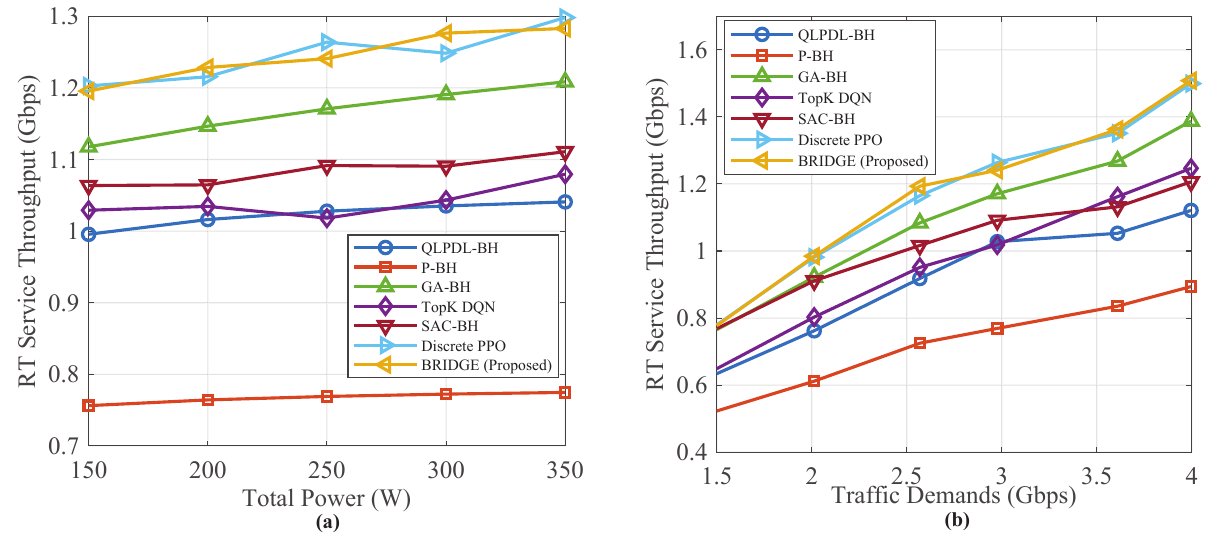}
\caption{RT service throughput of different algorithms. (a) RT Service Throughput versus Total Power (Traffic Demand = 3 Gbps), (b) RT Service Throughput versus Traffic Demands ($P_{max}$ = 250 W).}
\label{fig_7}
\end{figure*}
\begin{figure*}[!t]
\centering
\includegraphics[width=0.8\textwidth]{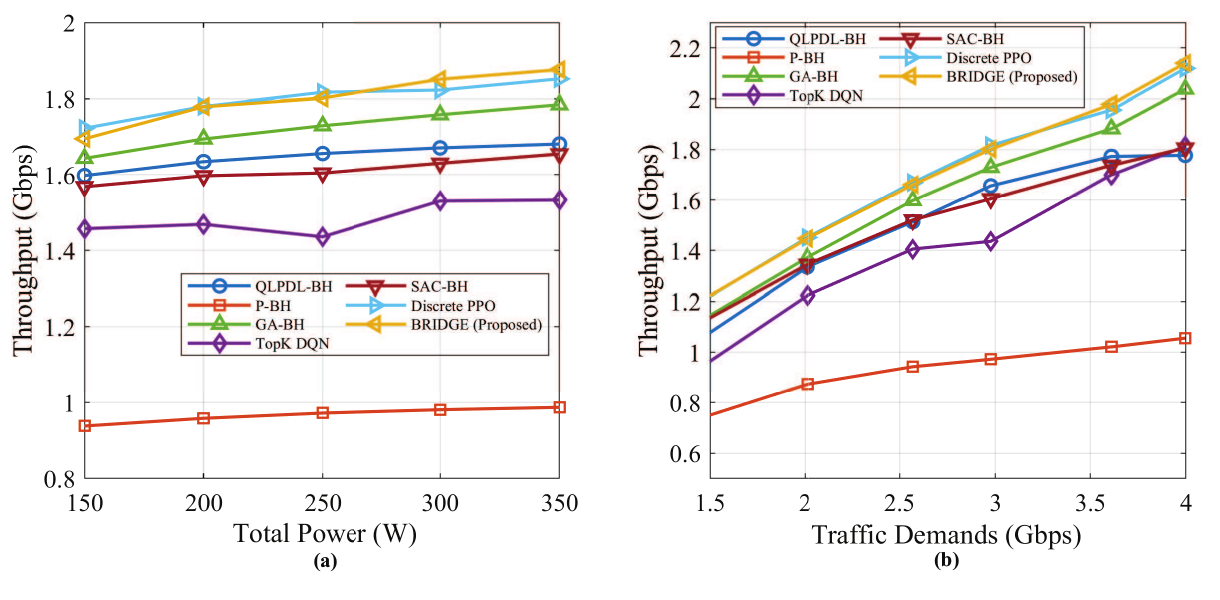}
\vspace{-4mm}
\caption{Throughput of different algorithms. (a) Throughput versus Total Power (Traffic Demand = 3 Gbps), (b) Throughput versus Traffic Demands ($P_{max}$ = 250 W).}
\label{fig_8}
\end{figure*} 

Fig. \ref{fig_7}–Fig. \ref{fig_8} present comparisons of RT service throughput and overall system throughput under the different conditions. The proposed BRIDGE method and its discrete variant, Discrete PPO, exhibit similar performance and consistently outperform the baseline algorithms, with their advantage becoming more pronounced in high-power and high-traffic scenarios. As total power increases, both RT service throughput and overall system throughput improve, although the growth gradually saturates for most algorithms. Meanwhile, due to on-board resource limitations and inter-beam frequency reuse constraints, QLPDL-BH stabilizes at approximately 1.78 Gbps when total traffic demand exceeds 3.5 Gbps. As illustrated in Fig.~\ref{fig_8}, the performance gap between GA-BH and QLPDL-BH in overall system throughput remains below 0.1 Gbps, while in RT service throughput, the difference rises to nearly 0.2 Gbps, as shown in Fig. \ref{fig_7}. These results indicate that, although some algorithms can partially maintain system throughput, they often fail to prioritize RT services. In contrast, BRIDGE and Discrete PPO not only preserve RT service performance but also enhance overall system throughput, thereby achieving a superior balance 
between 
the two.

\begin{figure*}[!t]
\centering
\includegraphics[width=0.9\textwidth]{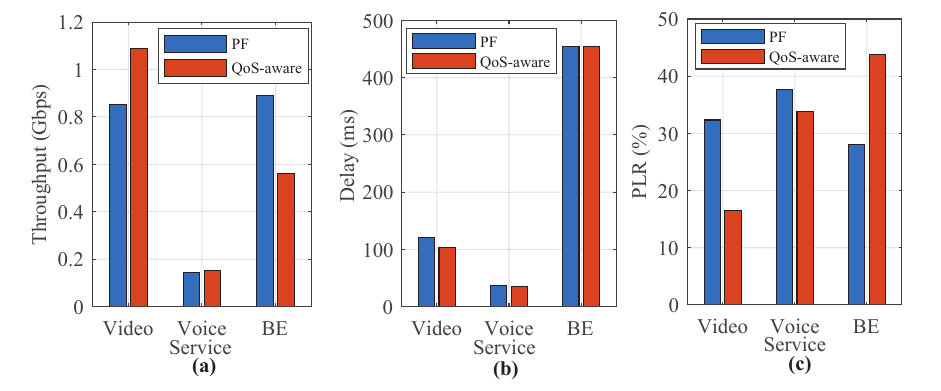}
\vspace{-4mm}
\caption{Comparison of QoS performance between PF-based and QoS-aware subchannel allocation. (a) Throughput, (b) Delay, and (c) PLR.}
\label{fig_13}
\end{figure*}

As shown in Fig.~\ref{fig_13}, the QoS performance of the proportional fair (PF)-based and QoS-aware subchannel allocation schemes, which are both built upon the BRIDGE framework for beam scheduling and power allocation, is evaluated under different service types. The results show that the QoS-aware scheme achieves higher throughput for Video and Voice services, whereas the PF-based scheme provides better performance for BE service. In terms of delay, the QoS-aware scheme effectively reduces the transmission delay of RT services. Moreover, the PLR of Video and Voice services are reduced by approximately 49\% and 10\%, respectively, at the expense of degraded BE performance. These results demonstrate that the proposed QoS-aware subchannel allocation scheme satisfies heterogeneous QoS requirements more effectively in multi-service scenarios.
\subsubsection{Performance of Fairness}
\begin{figure}[!t]
\centering
\includegraphics[width=0.9\columnwidth]{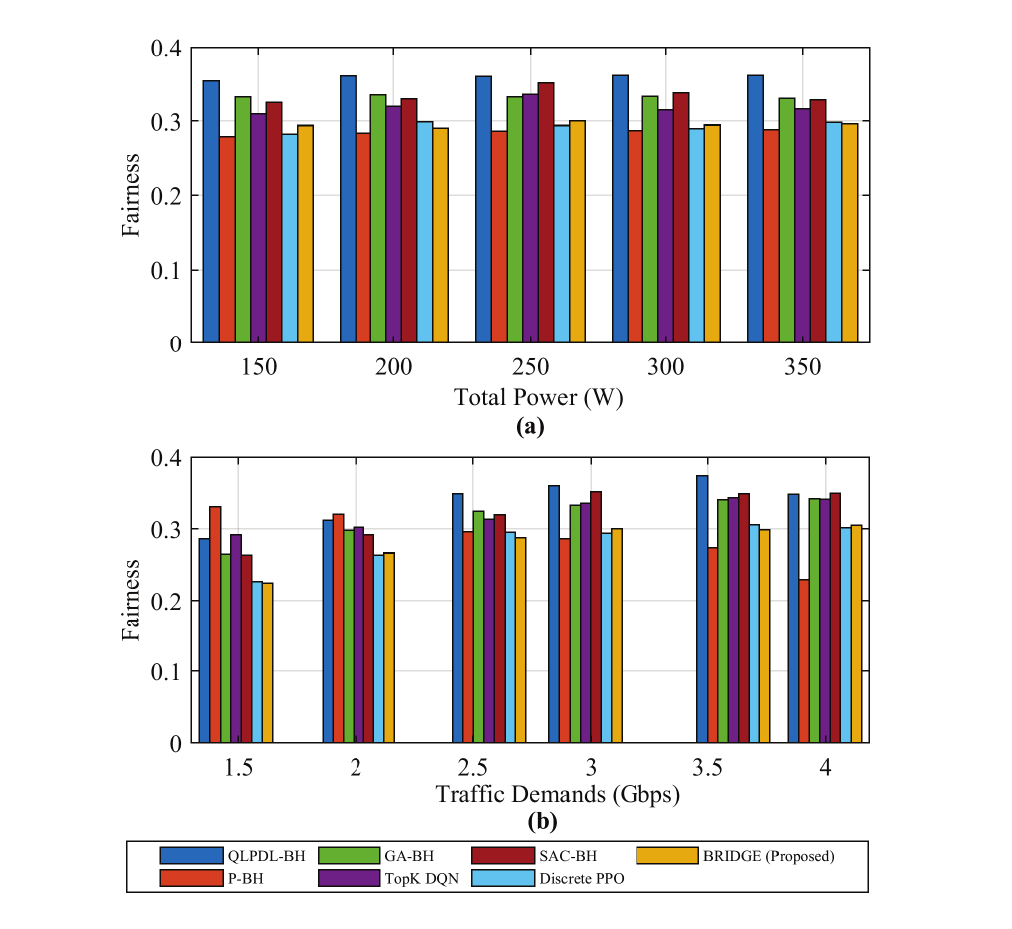}
\vspace{-4mm}
\caption{Fairness of different algorithms. (a) Fairness versus Total Power (Traffic Demand = 3 Gbps), (b) Fairness versus Traffic Demands ($P_{max}$ = 250 W).}
\label{fig_9}
\end{figure}

Fairness under varying total power and traffic demand is illustrated in Fig. \ref{fig_9}. The results show that the proposed BRIDGE method consistently achieves a relatively low fairness, indicating a more balanced allocation of system resources. In contrast, the fairness of the P-BH algorithm improves with increasing traffic demand: it performs poorly under low-demand conditions (e.g., approximately 0.325 at 1.5 Gbps), but in high-demand scenarios (e.g., decreasing to around 0.225 at 4 Gbps), it can even surpass some learning-based algorithms. The trend occurs because, as overall user service satisfaction declines, the system-level fairness metric tends to increase. By comparison, QLPDL-BH maintains relatively high fairness indices (typically above 0.3), reflecting weaker fairness performance. Meanwhile, GA-BH, Top-K DQN, and SAC-BH achieve better fairness, as their optimization objectives explicitly include the minimization of the standard deviation of user satisfaction.

\subsubsection{Performance Under Adversarial Attack}

\begin{figure}[htbp]
    \centering
    % 子图 (a)
    \subfloat[]{%
        \includegraphics[width=0.4\textwidth]{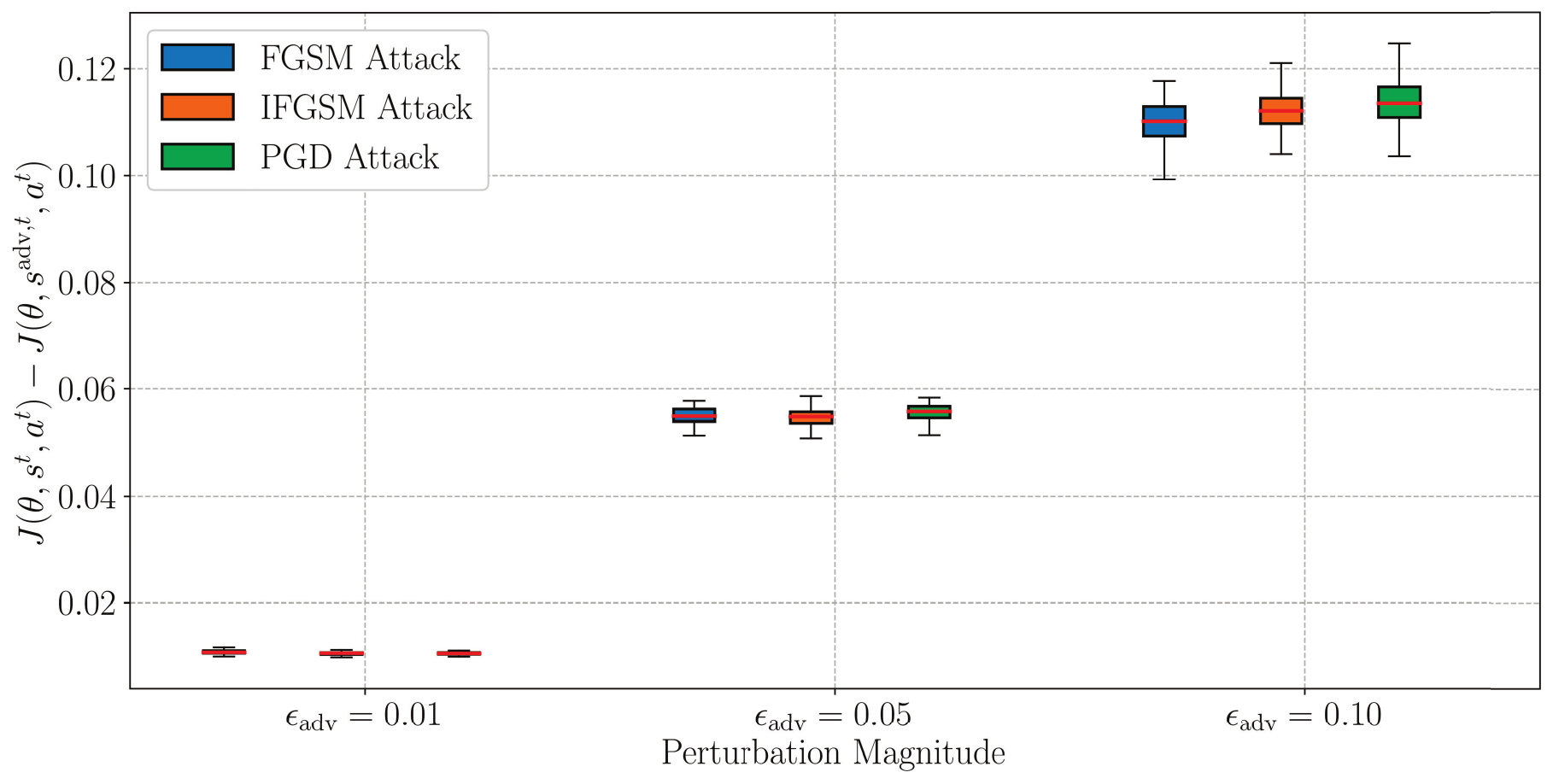}%
        \label{fig:subfig_a}%
    }
    \hfill
    % 子图 (b)
    \subfloat[]{%
        \includegraphics[width=0.4\textwidth]{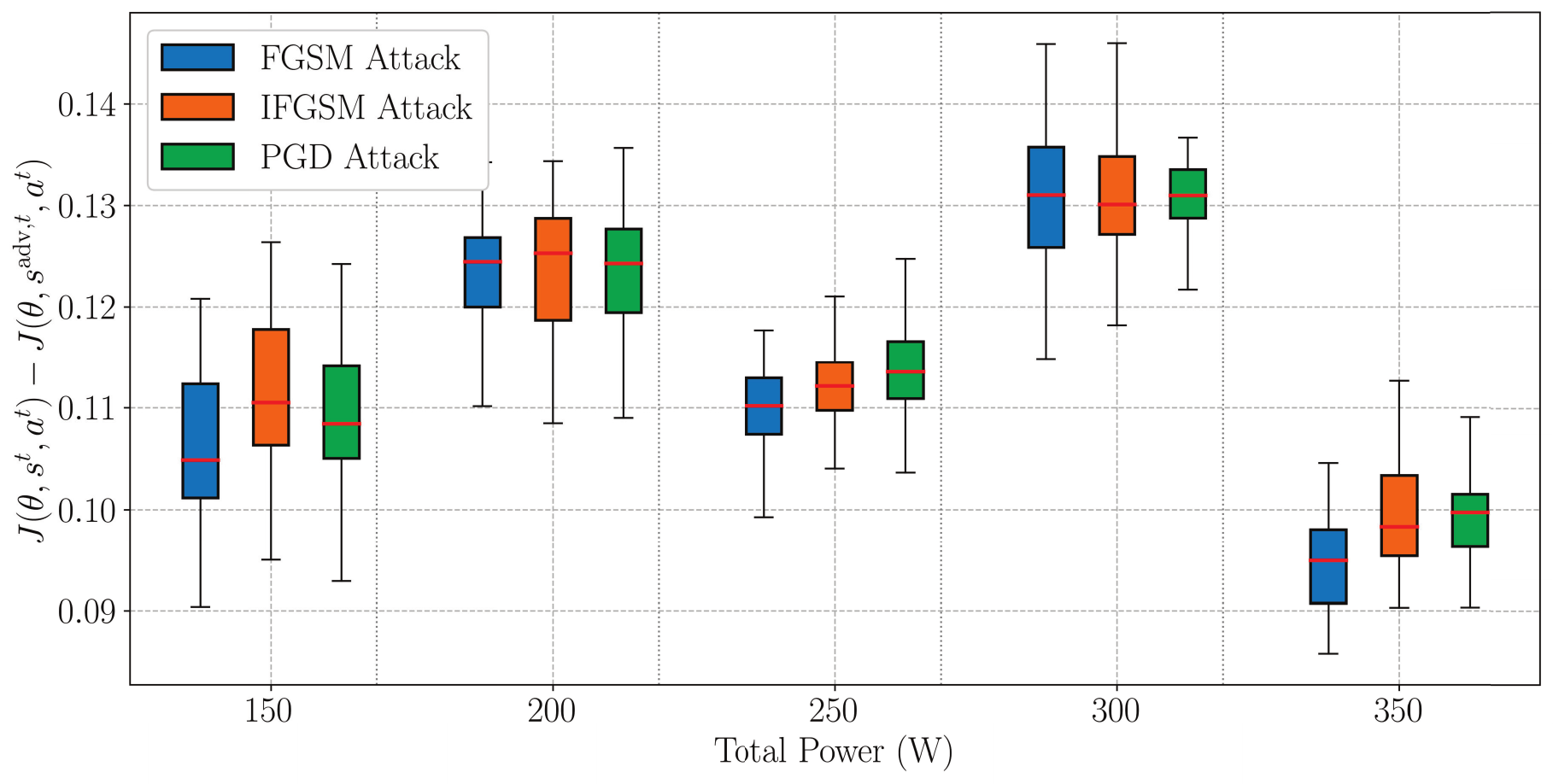}%
        \label{fig:subfig_b}%
    }
    \caption{Gradient deviation under adversarial attacks. 
    (a) Perturbation Magnitude (Traffic Demand = 3 Gbps and $P_{max}$ = 250 W); 
    (b) Total Power (Traffic Demand = 3 Gbps and $\epsilon_{adv}$ = 0.1).}
    \label{fig_10}
\end{figure}

\begin{figure}[!t]
\centering
\includegraphics[width=0.9\columnwidth]{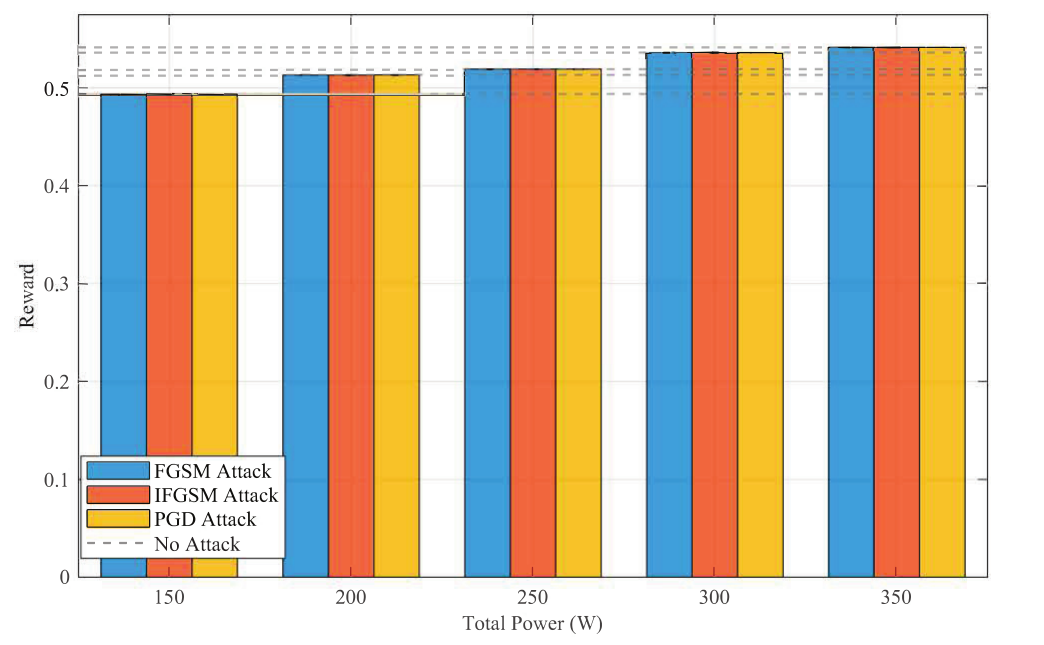}
\caption{Reward under adversarial attacks (Traffic Demand = 3 Gbps and $\epsilon_{adv}$ = 0.1).}
\label{fig_11}
\end{figure}

\begin{figure}[!t]
\centering
\includegraphics[width=\columnwidth]{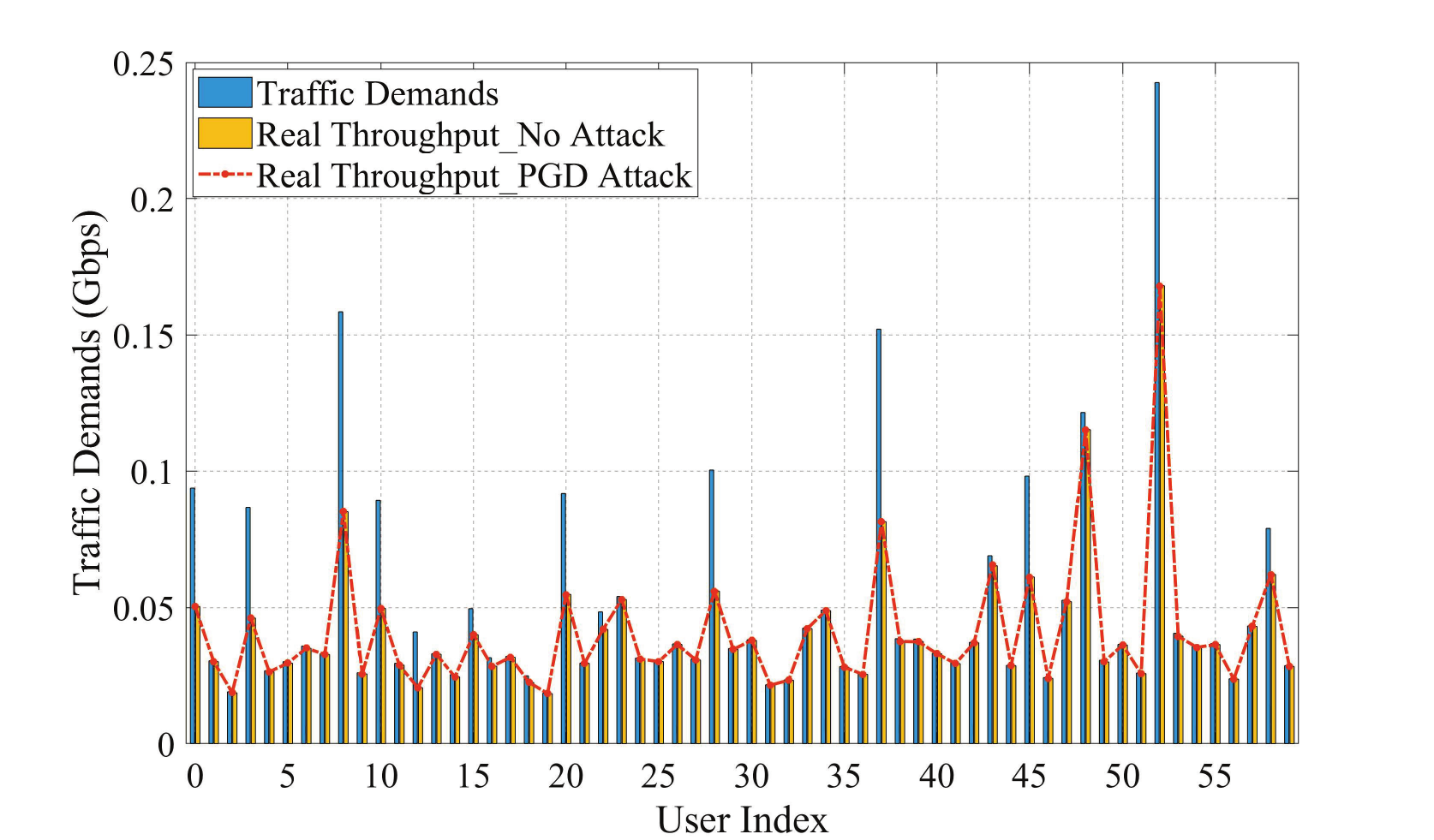}
\caption{The throughput of each user by the proposed BRIDGE algorithm with no attack and PGD attack ($\epsilon_{adv}=0.1$) under Traffic Demand = 1.5 Gbps and $P_{max}$ = 250 W.}
\label{fig_12}
\end{figure}

Fig.~\ref{fig_10}-\ref{fig_12} compare the performance of different adversarial attack methods. At low perturbation strengths, all three attacks exhibit only minor influence on the model. As the perturbation strength increases (see Fig. \ref{fig_10}(a)), the impact becomes more evident, with larger deviations in the cost function indicating stronger model disruption. Additional analysis under varying power conditions (see Fig. \ref{fig_10}(b)) shows that the absolute perturbation values across the three methods remain around 0.12 and exhibit similar trends: PGD and IFGSM produce stronger effects, whereas FGSM is consistently the weakest.

Fig.~\ref{fig_11} illustrates the reward performance of the model under different adversarial attacks with perturbation strength $\epsilon_{adv}=0.1$. The results indicate that, regardless of whether the model is subjected to FGSM, IFGSM, or the stronger PGD attack, its performance remains almost indistinguishable from the no-attack case. In other words, the rewards obtained during inference are nearly unaffected by adversarial perturbations. This suggests that, across different power settings, the considered adversarial attacks fail to substantially degrade model performance, confirming their limited effectiveness. Furthermore, as shown in Fig. \ref{fig_12}, when the total service demand is relatively low (1.5 Gbps), most users’ demands are nearly fully satisfied except for a few high-demand users. This highlights the capability of the proposed BRIDGE algorithm to effectively manage diverse traffic demands. Notably, even under PGD attacks, the model’s performance remains virtually unaffected.

In summary, the results show that BRIDGE maintains stable resource allocation performance under the bounded link gain perturbations generated by FGSM, I-FGSM, and PGD. This is mainly because the adopted attack model is restricted to a fixed perturbation budget and only perturbs the perceived link-gain component of the state. Since the policy also uses queue states, user positions, beam indices, and the remaining service time, the final scheduling decision is not determined by the link-gain information alone.

The design of BRIDGE also helps reduce the impact of such input disturbances. In particular, the Gumbel-TopK branch selects multiple beams according to their relative ranking, so a bounded perturbation changes the selected beam set only when it is large enough to alter the Top-K order of candidate beams. These results suggest that, although adversarial samples can affect AI-based scheduling models, their impact is limited under the fixed-budget perturbation setting considered in this work. As a result, BRIDGE maintains effective resource scheduling performance under the tested attacks.

\section{Conclusion}
\label{sec7}
This paper presents a joint optimization framework for beam scheduling and power allocation in BH-enabled LEO satellite systems by integrating the digital twin technology with deep reinforcement learning. More specifically, the proposed BH with reinforcement learning incorporating the integrated Dirichlet and Gumbel-TopK Exploration (BRIDGE) algorithm effectively handles hybrid discrete–continuous action spaces, incorporates QoS-driven subchannel scheduling, and achieves superior performance in terms of energy efficiency, throughput, and fairness. The robustness analysis shows that the AI-enabled scheduling policy maintains stable performance under the considered bounded  perturbations, supporting its applicability to future LEO satellite networks.

Future research will focus on validating the proposed framework using real-world satellite network data and more sophisticated channel models, as well as extending the proposed optimization mechanisms to large-scale satellite constellations. In particular, incorporating uncertainty-aware or probabilistic DT models to capture orbital prediction errors, environmental dynamics, and sensing uncertainties is an important direction. Such extensions would enable a more systematic investigation of uncertainty propagation and its impact on RL policy robustness. Furthermore, cross-layer security strategies will be explored to mitigate systemic cyber-physical threats. Ultimately, these efforts aim to support secure, resilient, and high-performance 6G space–air–ground integrated networks.

\bibliographystyle{IEEEtran}
\bibliography{mybib}
\vfill

\end{document}